\batchmode
\makeatletter
\def\input@path{{"C:/Trabajo laptop/Mis articulos/Finished/Tree PMBM spawning/Accepted/"}}
\makeatother
\documentclass[english]{IEEEtran}
\usepackage[T1]{fontenc}
\usepackage[latin9]{inputenc}
\usepackage{array}
\usepackage{float}
\usepackage{multirow}
\usepackage{amsmath}
\usepackage{amsthm}
\usepackage{amssymb}
\usepackage{graphicx}

\makeatletter

\providecommand{\tabularnewline}{\\}
\floatstyle{ruled}
\newfloat{algorithm}{tbp}{loa}
\providecommand{\algorithmname}{Algorithm}
\floatname{algorithm}{\protect\algorithmname}

\theoremstyle{plain}
\newtheorem{thm}{\protect\theoremname}
\theoremstyle{definition}
\newtheorem{defn}[thm]{\protect\definitionname}
\theoremstyle{definition}
\newtheorem{example}[thm]{\protect\examplename}
\theoremstyle{plain}
\newtheorem{prop}[thm]{\protect\propositionname}

\usepackage{cite}
\allowdisplaybreaks 

\usepackage[margin=8pt,font=footnotesize]{caption}
\usepackage{algorithm}
\usepackage{algpseudocode}
\usepackage{amsmath}  

\makeatother

\usepackage{babel}
\providecommand{\definitionname}{Definition}
\providecommand{\examplename}{Example}
\providecommand{\propositionname}{Proposition}
\providecommand{\theoremname}{Theorem}

\begin{document}
\title{Tracking multiple spawning targets using Poisson multi-Bernoulli mixtures
on sets of tree trajectories}
\author{Ángel F. García-Fernández, Lennart Svensson\thanks{A. F. García-Fernández is with the Department of Electrical Engineering and Electronics, University of Liverpool, Liverpool L69 3GJ, United Kingdom, and also with the ARIES Research Centre, Universidad Antonio de Nebrija,  Madrid, Spain (angel.garcia-fernandez@liverpool.ac.uk). L. Svensson is with the Department of Electrical Engineering, Chalmers University of Technology, SE-412 96 Gothenburg, Sweden (lennart.svensson@chalmers.se).} }
\maketitle
\begin{abstract}
This paper proposes a Poisson multi-Bernoulli mixture (PMBM) filter
on the space of sets of tree trajectories for multiple target tracking
with spawning targets. A tree trajectory contains all trajectory information
of a target and its descendants, which appear due to the spawning
process. Each tree contains a set of branches, where each branch has
trajectory information of a target or one of the descendants and its
genealogy. For the standard dynamic and measurement models with multi-Bernoulli
spawning, the posterior is a PMBM density, with each Bernoulli having
information on a potential tree trajectory. To enable a computationally
efficient implementation, we derive an approximate PMBM filter in
which each Bernoulli tree trajectory has multi-Bernoulli branches,
obtained by minimising the Kullback-Leibler divergence. The resulting
filter improves tracking performance of state-of-the-art algorithms
in a simulated scenario.
\end{abstract}


\section{Introduction}

The main goal of multiple target tracking (MTT) is to estimate the
trajectories of an unknown number of targets that may appear, move
and disappear using noisy and clutter measurements \cite{Blackman_book99}.
Applications can be found in sensor networks \cite{Gao19}, autonomous
vehicles \cite{Chen19}, and cell biology \cite{Chenouard14,Magnusson15}.

In Bayesian MTT, the standard multi-target dynamic model considers
probabilistic models for new born targets, single target dynamics
and death events \cite{Mahler_book14,Meyer18}. In this setting, all
information about the target trajectories is encapsulated in the density
of the set of trajectories given past and current measurements, which
is referred to as the posterior density \cite{Angel20_b,Coraluppi14}.
For the standard measurement model, if the birth process is a Poisson
point process (PPP), the posterior density on the set of trajectories,
and also on the current set of targets, is a Poisson multi-Bernoulli
mixture (PMBM) \cite{Granstrom18,Williams15b,Angel18_b}. The PMBM
density models information on undetected targets/trajectories with
a PPP, and information on detected targets/trajectories with a multi-Bernoulli
mixture (MBM). The PMBM posterior becomes an MBM if the birth process
is multi-Bernoulli \cite{Xia19_b}. 

In some applications, there may be targets that are spawned from other
targets \cite{Isaac08}, for instance, a cell may undergo mitosis
(cell division) \cite{Dzyubachyk10}, a skydiver may jump from an
airplane, or a spacecraft may break up \cite{Flegel17}. This possibility
can be included in the dynamic model by adding a spawning model \cite{Mahler_book14,Granstrom13b}.
Several filters have been developed to estimate the current set of
targets with spawning, for example, the probability hypothesis density
(PHD) filter \cite{Mahler03,Mahler_book14}, the cardinality PHD (CPHD)
filter \cite{Lundgren13,Bryant17}, the PMBM filter \cite{Su21} and
the generalised labelled multi-Bernoulli (GLMB) filter \cite{Bryant18,Nguyen21,Xu21},
which provides genealogy information on the targets. 

In this work, we show how to compute and approximate a posterior distribution
that contains full trajectory and genealogy information for all targets,
which none of the above methods provide. We consider a PPP birth model
and a multi-Bernoulli spawning model, in which each alive target can
spawn a finite number of targets. We first define the space of tree
trajectories, which enables us to represent all the information related
to a target trajectory and its different branches arising from spawning.
Trees have been used to model the genealogy of branching processes
\cite{Shi_book15,Popovic04}.  Here, we define a tree trajectory
by a start time and a set of branches, each with its genealogy and
sequence of states (trajectory). A set of tree trajectories can therefore
represent all target trajectories and their genealogies. 

\begin{figure}
\begin{centering}
\includegraphics[scale=0.6]{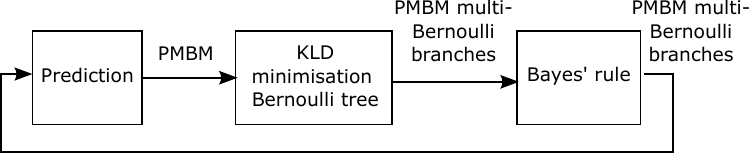}
\par\end{centering}
\caption{\label{fig:Diagram}Diagram of the proposed PMBM recursion for sets
of tree trajectories with multi-Bernoulli branches in each Bernoulli
tree trajectory. After the prediction step, we perform a KLD minimisation
on each Bernoulli tree trajectory density to obtain multi-Bernoulli
branches.}
\end{figure}

In this paper, we show that the posterior density on the set of tree
trajectories is a PMBM, making use of the PMBM update for generalised
measurements \cite{Angel21}. The resulting filter is referred to
as the tree PMBM (TrPMBM) filter, and the tree MBM (TrMBM) filter
for multi-Bernoulli birth. In this setting, each Bernoulli density
in the TrPMBM has information on a potential tree trajectory. We also
provide an efficient implementation of the TrPMBM filter by making
a multi-Bernoulli branch approximation in each Bernoulli tree trajectory,
and redefining the global hypothesis at a branch level, instead of
at a tree level. The multi-Bernoulli branch approximation is obtained
by minimising the Kullback-Leibler divergence (KLD) after each prediction
step, see Figure \ref{fig:Diagram} for a diagram. Finally, we propose
an implementation of the TrPMBM filter for linear/Gaussian models
and evaluate the results via numerical simulations.

The rest of the paper is organised as follows. The problem formulation
is presented in Section \ref{sec:Problem-formulation}. The dynamic
and measurement models for tree trajectories are provided in Section
\ref{sec:Dynamic/measurement-models-tree}. The TrPMBM posterior and
its approximation with multi-Bernoulli branches are explained in Section
\ref{sec:TrPMBM_filter}. The TrPMBM filter recursion is given in
Section \ref{sec:TrPMBM-filter-recursion} and its Gaussian implementation
in Section \ref{sec:Gaussian-TrPMBM-filter}. Simulation results and
conclusions are provided in Sections \ref{sec:Simulations} and \ref{sec:Conclusions},
respectively.

\section{Problem formulation\label{sec:Problem-formulation}}

We aim to obtain the posterior density of the set of all tree trajectories.
This density contains all information on target trajectories that
have ever entered the surveillance area and their genealogies arising
from spawning processes. The dynamic and measurement models for targets
is described in Section \ref{subsec:Models_targets}. The space of
sets of tree trajectories is defined in Section \ref{subsec:Space-tree-trajectories}.
Integration for sets of tree trajectories is explained in Section
\ref{subsec:Integration}. The main notation of this paper is summarised
in Table \ref{tab:Notation}. 

\begin{table}
\caption{\label{tab:Notation}Notation}

\rule[0.5ex]{1\columnwidth}{1pt}
\begin{itemize}
\item $\mathbf{x}_{k}$: set of targets at time step $k$, $x\in\mathbf{x}_{k}$
is a target state.
\item $\mathbf{X}_{k}$: set of all tree trajectories up to time step $k$.
\item $X=\left(t,\mathbf{B}\right)\in\mathbf{X}_{k}$: a tree trajectory
with start time $t$ and set $\mathbf{B}$ of branches.
\item $B=\left(\omega,x^{1:\ell\left(\omega\right)}\right)$: a branch with
genealogy variable $\omega$, length $\ell\left(\omega\right)$ and
states $x^{1:\ell\left(\omega\right)}$.
\item $\omega=\left(\omega^{1},...,\omega^{\nu}\right)$: genealogy variable
of a branch in a tree with at most $\nu$ generations.
\item $\omega_{(\nu,j,l)}$: genealogy variable of the $j$-th branch in
a tree with at most $\nu$ generations, and branch length $l$.
\item $\mathbf{B}^{j}$: set with the $j$-th branch in a tree.
\item $p_{k|k'}^{i,j,\alpha}\left(\mathbf{B}^{j}\right)$: Bernoulli density
of the $j$-th branch in tree $i$ under global branch hypothesis
$\alpha$, at time step $k$ given measurements up to time step $k'$
with
\begin{itemize}
\item Single branch density $p_{k|k'}^{i,j,\alpha}\left(B\right)$ and probability
$r_{k|k'}^{i,j,\alpha}$ of existence.
\item $\beta_{k|k'}^{i,j,\alpha}\left(\kappa\right)$: probability that
the branch ends at time step $\kappa$.
\item $\overline{\omega}_{k,\kappa}^{i,j}$: genealogy variable with the
branch ending at time step $\kappa$.
\item $\overline{x}_{k|k'}^{i,j,\alpha}\left(\kappa\right)$: mean given
that the branch ends at time step $\kappa$.
\item $P_{k|k'}^{i,j,\alpha}\left(\kappa\right)$: covariance given that
the branch ends at time step $\kappa$.
\end{itemize}
\item $w_{k|k'}^{i,j,\alpha}$: weight of the $j$-th branch in the $i$-th
tree in global branch hypothesis $\alpha$ at time step $k$ given
measurements up to time step $k'$.
\item $1_{\ell}$: sequence of length $\ell$ with all ones.
\end{itemize}
\rule[0.5ex]{1\columnwidth}{1pt}
\end{table}

\subsection{Dynamic and measurements models for targets\label{subsec:Models_targets}}

We denote a target state as $x\in\mathbb{R}^{n_{x}}$ and a measurement
state as $z\in\mathbb{R}^{n_{z}}$. Let $\mathbf{x}_{k}$ and $\mathbf{z}_{k}$
be the sets of targets and measurements at time step $k$, respectively.
The dynamic model is characterised as follows. At each time step:
\begin{itemize}
\item Each target $x\in\mathbf{x}_{k}$ may survive to the next time step
with survival probability $p_{1}^{S}\left(x\right)$ and transition
density $g_{1}\left(\cdot|x\right)$. 
\item Each target $x\in\mathbf{x}_{k}$ may spawn with $\varrho-1$ independent
modes with spawning probability $p_{m}^{S}\left(x\right)$ and transition
density $g_{m}\left(\cdot|x\right)$ for $m\in\{2,...,\varrho\}$.
\item The birth model is a PPP with intensity $\lambda^{B}\left(\cdot\right)$.
\end{itemize}
The standard measurement model is:
\begin{itemize}
\item Each target $x\in\mathbf{x}_{k}$ is detected with probability $p^{D}\left(x\right)$
and generates a measurement with density $l\left(\cdot|x\right)$. 
\item Clutter is a PPP with intensity $\lambda^{C}\left(\cdot\right)$.
\end{itemize}
The measurement set $\mathbf{z}_{k}$ is the union of target-generated
measurements and clutter at time step $k$.

It should be noted that, given a target state $x$, the distribution
of the set of targets that survive or spawn at the next time step
is multi-Bernoulli with existence probabilities and single target
densities $\left\{ \left(p_{1}^{S}\left(x\right),g_{1}\left(\cdot|x\right)\right),...,\left(p_{\varrho}^{S}\left(x\right),g_{\varrho}\left(\cdot|x\right)\right)\right\} $.
As targets usually survive with high probability and spawn with low
probability, $p_{1}^{S}\left(x\right)$ is high and $p_{m}^{S}\left(x\right)$
for $m\geq2$ is low. 

\subsection{Space of tree trajectories\label{subsec:Space-tree-trajectories}}

\begin{figure}
\begin{centering}
\includegraphics[scale=0.6]{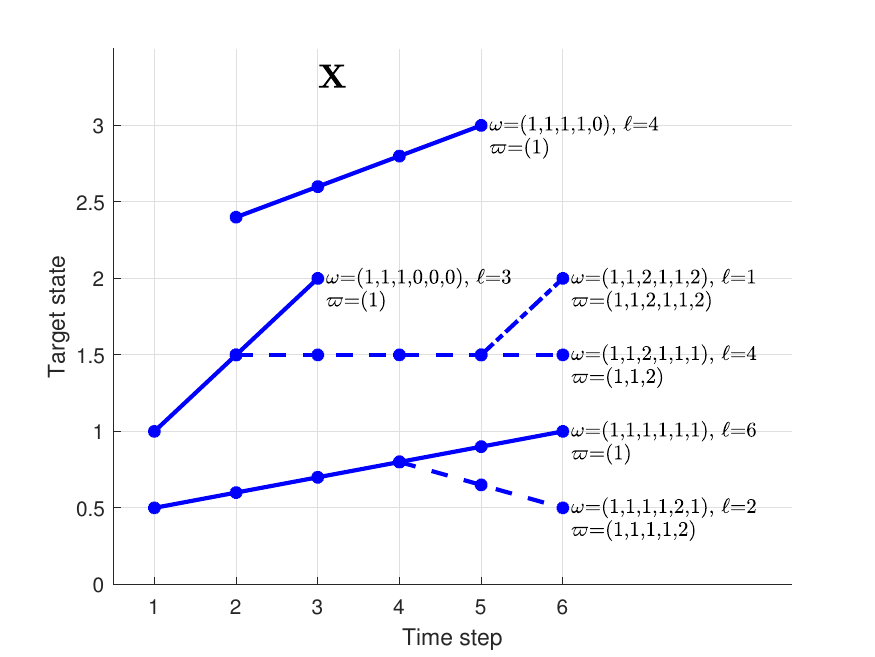}
\par\end{centering}
\caption{\label{fig:Illustration-tree_trajectory}Illustration of a set of
tree trajectories, with three tree trajectories, $\mathbf{X}=\left\{ X_{1},X_{2},X_{3}\right\} $.
The main branch (surviving mode) is represented as a continuous line.
The dashed lines represent descendant branches from a main branch
and the dashed dotted line a descendant branch from a secondary branch.
At the end of each branch, it is written its genealogy variable $\omega$
up to time step 6, its ID $\varpi$ and its length $\ell$.}
\end{figure}

With target spawning, each target born at a given time initiates a
genealogy of surviving/spawned targets at the following time steps.
All information regarding the genealogies and the trajectories of
a target and its descendants is included in a tree trajectory. A tree
trajectory has a main branch, corresponding to the survival of the
target that originated from the birth process. New branches appear
for each spawning from this target or one of its descendants. The
information on all targets that have been present in the surveillance
area, their trajectories and genealogies can then be represented by
a set of tree trajectories, see Figure \ref{fig:Illustration-tree_trajectory}
for an illustration. We proceed to describe the spaces of single tree
trajectories and of sets of tree trajectories.

Let $t$ denote the time step when a target is born. At each of the
following time steps $t+i-1$, $i>1$, this target and its descendants
may generate new targets according to the spawning process. We refer
to the time index $i$ as the $i$-th generation of the tree. The
maximum number of generations in a tree is denoted by $\nu$. For
example, if the current time step is $k$, $\nu=k-t+1$. Each spawning
process from this target or its descendants generates a new branch. 
\begin{defn}
\label{def:Genealogy} Given a branch that spawned at generation
$i$ in a tree with at most $\nu$ generations, its genealogy $\omega=\left(\omega^{1},...,\omega^{\nu}\right)$
is defined as follows. Variable $\omega^{i}\in\left\{ 2,...,\varrho\right\} $
contains its spawning mode. For subsequent generations $j>i$, $\omega^{j}=1$
or $\omega^{j}=0$ if the branch is present or not present at generation
$j$, respectively. For generations previous to the branch spawning,
$j<i$, the branch inherits the genealogy variable $\omega^{j}$ of
the parent branch. In particular, $\omega^{1}=1$ by convention indicating
that the main branch of a tree is created by target birth. 
\end{defn}
The space of the genealogy variables is denoted by $I_{(\nu)}$ so
that $\omega\in I_{(\nu)}$.
\begin{defn}
A branch $B=\left(\omega,x^{1:\ell\left(\omega\right)}\right)$ consists
of the genealogy variable $\omega$ and the sequence of states $x^{1:\ell\left(\omega\right)}=\left(x^{1},...,x^{\ell\left(\omega\right)}\right)$,
where $x^{1}$ is the state at the time when the branch is spawned
or born and $x^{2:\ell\left(\omega\right)}$ is the sequence of states
at the following time steps. The length $\ell\left(\omega\right)$
of the branch is the number of time steps that the branch is present,
which is a deterministic function of $\omega$, see Appendix \ref{subsec:Branch-length}. 
\end{defn}
 The single branch space in tree with at most $\nu$ generations
is then $\mathbb{B}_{(\nu)}=\uplus_{\omega\in I_{(\nu)}}\left\{ \omega\right\} \times\mathbb{R}^{\ell\left(\omega\right)\cdot n_{x}}$,
where $\uplus$ denotes the union of disjoint sets \cite{Mahler_book14}.
That is, $\mathbb{B}_{(\nu)}$ contains all possible genealogy variables
$\omega$, to which we append a sequence of target states of suitable
length. 
\begin{defn}
A tree trajectory $X$ is a variable formed by the tree starting time
$t$ and its set $\mathbf{B}$ of branches, such that $X=\left(t,\mathbf{B}\right)$
with $\mathbf{B}=\left\{ B_{1},...,B_{n}\right\} $. 
\end{defn}
The space of single tree trajectories from time step 1 to $k$ is
\begin{align}
\mathbb{T}_{(k)} & =\uplus_{t=1}^{k}\left[\left\{ t\right\} \times\mathcal{F}\left(\mathbb{B}_{(k-t+1)}\right)\right]\label{eq:tree_trajectory_space_1_k}
\end{align}
where $\mathcal{F}\left(\mathbb{B}_{(\nu)}\right)$ is the set of
all finite subsets of $\mathbb{B}_{(\nu)}$. That is, the starting
time $t$ belongs to $\{1,...,k\}$ and, for a given $t$, the branches
have at most $\nu=k-t+1$ generations. Therefore, the set of branches
belongs to $\mathcal{F}\left(\mathbb{B}_{(k-t+1)}\right)$. In addition,
the genealogy variables of the branches in a tree require several
constraints, described in Appendix \ref{subsec:Genealogy-constraints},
so that an element of (\ref{eq:tree_trajectory_space_1_k}) corresponds
to a tree. These constraints are ensured by a suitable birth and dynamic
model, as in target labelling approaches, where label uniqueness is
ensured probabilistically \cite{Bryant18}. 

We should note that, according to the spawning process in Section
\ref{subsec:Models_targets}, the maximum number of branches in a
tree trajectory with at most $\nu$ generations is

\begin{equation}
n_{max}\left(\nu\right)=\varrho^{\nu-1}.\label{eq:Maximum_number_branches}
\end{equation}

\begin{defn}
\label{def:Branch-ID}The ID $\varpi$ of a branch in a tree $X=\left(t,\mathbf{B}\right)$
is the branch genealogy variable $\omega$ up to the generation when
the branch was spawned or born. 
\end{defn}
The ID $\varpi$ uniquely identifies each branch in a tree as there
cannot be more than one branch in a tree with the same $\varpi$.
Finally, a set of tree trajectories is denoted by $\mathbf{X}\in\mathcal{F}\left(\mathbb{T}_{(k)}\right)$.
We illustrate the previous concepts via the following example.
\begin{example}
\label{exa:Set_trees}We consider the set $\mathbf{X}$ of tree trajectories
in Figure \ref{fig:Illustration-tree_trajectory}. The tree trajectory
in the middle has start time $t=1$, $\nu=6$ and three branches,
each with the genealogy variables, branch ID and lengths shown in
the figure. The branch with $\omega=\left(1,1,2,1,1,2\right)$ was
spawned at $\nu=6$, and has length 1. Its genealogy indicates that
its parent branch spawned from the main branch at generation 3, as
there is a 2 in the third entry in $\omega$. The tree trajectory
at the top has start time $t=2$, $\nu=5$ and only has the main branch.
The last entry in its genealogy is zero because it is not alive at
time step 6. 
\end{example}

\subsection{Integration\label{subsec:Integration}}

The space of single tree trajectories is locally compact, Hausdorff
and second-countable (LCHS), see Appendix \ref{subsec:Tree-trajectory-LCHS}.
Then, we can use finite set statistics to define probabilistic models
and integrals for sets of tree trajectories \cite{Mahler_book14}.
Given a real-valued function $\pi\left(\cdot\right)$ on $\mathbb{T}_{(k)}$,
its single tree integral is
\begin{align}
\int\pi\left(X\right)dX & =\sum_{t=1}^{k}\int_{\mathbb{B}_{(k-t+1)}}\pi\left(t,\mathbf{B}\right)\delta\mathbf{B}\label{eq:tree_trajectory_integral_general}
\end{align}
where we first sum over all possible start times, and the set integral
is on the space of the set of branches with start time $t$ and maximum
end time $k$. The set integral in (\ref{eq:tree_trajectory_integral_general})
can also be written explicitly, see Appendix \ref{subsec:App_explicit-single-tree-integral}.
A function $\pi\left(\cdot\right)$ represents a density of a tree
trajectory if $\pi\left(\cdot\right)\geq0$ and it integrates to one.
To be precise, this density should also ensure that the genealogy
variables are unique and meet the genealogy constraints (parent/children
relations), see Appendix \ref{subsec:Genealogy-constraints}. 

Given a function $\pi\left(\cdot\right)$ on $\mathcal{F}\left(\mathbb{T}_{(k)}\right)$,
its set integral is \cite{Mahler_book14} 
\begin{align}
\int\pi\left(\mathbf{X}\right)\delta\mathbf{X} & =\sum_{n=0}^{\infty}\frac{1}{n!}\int\pi\left(\left\{ X_{1},...,X_{n}\right\} \right)dX_{1:n}.\label{eq:Set_integral}
\end{align}
A function $\pi\left(\cdot\right)$ represents a density of a set
of tree trajectories if $\pi\left(\cdot\right)\geq0$, it integrates
to one, and each tree meets the genealogy constraints. Equation (\ref{eq:Set_integral})
is a nested set integral, as each single tree trajectory integral
has a set integral. Nested set integrals have been used for group
targets \cite{Mahler01,Swain10}\cite[Chap. 21]{Mahler_book14}. 

\subsubsection{Set of branches integration\label{subsec:Alternative-branch-integration}}

The single tree trajectory integral (\ref{eq:tree_trajectory_integral_general})
requires the calculation of an integral over the set of branches with
$\nu=k-t+1$ generations. In this section, we explain a way to perform
this set integral that will be useful to derive the PMBM filter in
Section \ref{sec:TrPMBM_filter}.

The number of unique branch IDs, see Definition \ref{def:Branch-ID},
in a set of branches $\mathbf{B}\in\mathcal{F}\left(\mathbb{B}_{(\nu)}\right)$
is $n_{max}\left(\nu\right)$, see (\ref{eq:Maximum_number_branches}).
We arrange the branch IDs in a tree in lexicographical order (or any
other order) to refer to each branch by an index $j=1,...,n_{max}\left(\nu\right)$.
In a tree with $\nu$ generations, the $j$-th branch can have lengths
from 1 to $\ell_{max}\left(\nu,j\right)$, which is used to denote
the number of generations since its birth/spawning to $\nu$. Given
$\nu$ and the branch length $l$, we can obtain the genealogy variable
of the $j$-th branch, $\omega_{(\nu,j,l)}$. 

We have that $\mathbb{B}_{(\nu)}=\uplus_{j=1}^{n_{max}\left(\nu\right)}\mathbb{B}_{(\nu,j)}$
where $\mathbb{B}_{(\nu,j)}$ is the space of branch $j$ in a tree
with at most $\nu$ generations, which can be written as $\mathbb{B}_{(\nu,j)}=\uplus_{\ell=1}^{\ell_{max}\left(\nu,j\right)}\left\{ \omega_{(\nu,j,l)}\right\} \times\mathbb{R}^{\ell\cdot n_{x}}$.
That is, the single branch space $\mathbb{B}_{(\nu)}$ is the union
of the spaces for each branch, which are disjoint sets as there cannot
be two branches in a tree with the same genealogy variable. Then,
we can decompose a given $\mathbf{B}$ as $\mathbf{B}^{1}\uplus...\uplus\mathbf{B}^{n_{max}\left(\nu\right)}=\mathbf{B}$,
where $\mathbf{B}^{j}$ is the set representing the $j$-th branch,
which meets $\left|\mathbf{B}^{j}\right|\leq1$. Due to this decomposition
of the single branch space into disjoint spaces, the set integral
for a function $\pi\left(\cdot\right)$ on $\mathcal{F}\left(\mathbb{B}_{(\nu)}\right)$
can be written as \cite[Sec. 3.5.3]{Mahler_book14}
\begin{align}
 & \int_{\mathbb{B}_{(\nu)}}\pi\left(\mathbf{B}\right)\delta\mathbf{B}\nonumber \\
 & =\int...\int\pi\left(\mathbf{B}^{1}\uplus...\uplus\mathbf{B}^{n_{max}\left(\nu\right)}\right)\delta\mathbf{B}^{1}...\delta\mathbf{B}^{n_{max}\left(\nu\right)}.
\end{align}
In addition, if $\pi\left(\cdot\right)$ is a density on a set of
branches, which implies $\pi\left(\mathbf{B}^{j}\right)=0$ for $\left|\mathbf{B}^{j}\right|>1$,
the set integral on $\mathbf{B}^{j}$ is
\begin{align}
 & \int\pi\left(\mathbf{B}^{j}\right)\delta\mathbf{B}^{j}\nonumber \\
 & =\pi\left(\emptyset\right)+\sum_{\ell=1}^{\ell_{max}\left(\nu,j\right)}\int\pi\left(\left\{ \left(\omega_{(\nu,j,\ell)},x^{1:\ell}\right)\right\} \right)dx^{1:\ell}.\label{eq:Bernoulli_branch_integral}
\end{align}
That is, the set integral for the (uniquely identified) potential
$j$-th branch goes through the hypotheses that it does not exist
and that it exists with a length $\ell$, where $\ell$ takes values
from 1 to its maximum possible length $\ell_{max}\left(\nu,j\right)$.
The corresponding genealogy variable is $\omega_{(\nu,j,\ell)}$.

\section{Dynamic/measurement models for tree trajectories\label{sec:Dynamic/measurement-models-tree}}

In this section, we write the single tree trajectory dynamic and measurement
models that arise from the single target models in Section \ref{subsec:Models_targets}
for computing the posterior density on the set of all tree trajectories. 

\subsection{Dynamic model for all tree trajectories\label{subsec:Dynamic-model-tree-trajectories}}

A tree trajectory up to time step $k-1$ can be written as $X_{k-1}=\left(t_{k-1},\mathbf{B}_{k-1}\right)$,
with the set of branches $\mathbf{B}_{k-1}=\mathbf{B}_{k-1}^{1}\uplus...\uplus\mathbf{B}_{k-1}^{n_{max}\left(k-t_{k-1}+1\right)}$,
where $\mathbf{B}_{k-1}^{j}$ is the set representing the $j$-th
branch, see Section \ref{subsec:Alternative-branch-integration}.
Each tree trajectory evolves independently of the rest and the dynamic
model is characterised by the probability of survival of a tree trajectory
$p^{S}(X_{k-1})$, single tree trajectory dynamic model $g\left(X_{k}|X_{k-1}\right)$
and PPP birth intensity $\lambda_{k}^{B}\left(X_{k}\right)$. 

We consider the dynamic model for all tree trajectories, which implies
that $p^{S}(X_{k-1})=1$, as all trajectories remain in the set of
interest even if a tree trajectory is no longer present at the current
time step \cite{Granstrom18,Angel20_e}. We denote the $m$-th branch
spawned from $\mathbf{B}_{k-1}^{j}$ by $\mathbf{B}_{k}^{j,m}$, being
$\mathbf{B}_{k}^{j,1}$ the surviving branch. As branches spawn independently,
the single tree transition density is
\begin{align}
 & g\left(t_{k},\uplus_{j,m}\mathbf{B}_{k}^{j,m}\left|t_{k-1},\mathbf{B}_{k-1}\right.\right)\nonumber \\
 & =\delta_{t_{k-1}}\left[t_{k}\right]\prod_{j=1}^{n_{max}\left(k-t_{k-1}\right)}\prod_{m=1}^{\varrho}g_{m}\left(\mathbf{B}_{k}^{j,m}\left|t_{k-1},\mathbf{B}_{k-1}^{j}\right.\right)\label{eq:transition_density_tree_all}
\end{align}
where $\delta_{t}\left[\cdot\right]$ is a Kronecker delta at $t$.
It should be noted that, in general, the density of a set that is
the union of independent sets contains a convolution sum \cite{Mahler_book14,Angel18_b}.
In (\ref{eq:transition_density_tree_all}), there is no convolution
sum as the unique identifiers in the genealogy variables provide the
right association between the branches at time step $k$ and their
parent at time step $k-1$.

If branch $\mathbf{B}_{k-1}^{j}=\emptyset,$ we have
\begin{align}
g_{m}\left(\mathbf{B}_{k}^{j,m}\left|t_{k-1},\emptyset\right.\right) & =\delta_{\emptyset}\left(\mathbf{B}_{k}^{j,m}\right)
\end{align}
for all $m$, where $\delta_{\emptyset}\left(\cdot\right)$ is a multi-object
Dirac delta evaluated at $\emptyset$ \cite{Mahler_book14}. That
is, if branch $j$ does not exist, its descendants do not exist.

If branch $\mathbf{B}_{k-1}^{j}=\left\{ \left(\omega,x^{1:\ell}\right)\right\} $
is not present at time step $k-1$ (the last element of $\omega$
is zero), then
\begin{align}
 & g_{1}\left(\mathbf{B}_{k}^{j,1}\left|t_{k-1},\left\{ \left(\omega,x^{1:\ell}\right)\right\} \right.\right)\nonumber \\
 & =\begin{cases}
\delta_{x^{1:\ell}}\left(y^{1:\ell}\right) & \mathbf{B}_{k}^{j,1}=\left\{ \left(\left(\omega,0\right),y^{1:\ell}\right)\right\} \\
0 & \mathrm{otherwise}
\end{cases}\label{eq:g_1_prediction_dead}
\end{align}
where $\delta_{x}\left(\cdot\right)$ denotes a Dirac delta at $x$
and
\begin{align}
g_{m}\left(\mathbf{B}_{k}^{j,m}\left|t_{k-1},\mathbf{B}_{k-1}^{j}\right.\right) & =\delta_{\emptyset}\left(\mathbf{B}_{k}^{j,m}\right)
\end{align}
for $m>1$. That is, we append a zero entry to $\omega$ and keep
the past trajectories information in the main branch, and there are
no spawning branches. 

If branch $\mathbf{B}_{k-1}^{j}=\left\{ \left(\omega,x^{1:\ell}\right)\right\} $
is present at time step $k-1$ (the last element of $\omega$ is higher
than zero), then, for $m=1$,
\begin{align}
 & g_{1}\left(\mathbf{B}_{k}^{j,1}\left|t_{k-1},\left\{ \left(\omega,x^{1:\ell}\right)\right\} \right.\right)\nonumber \\
 & =\begin{cases}
p_{1}^{S}\left(x^{\ell}\right)g_{1}\left(y^{\ell+1}|x^{\ell}\right)\\
\:\times\delta_{x^{1:\ell}}\left(y^{1:\ell}\right) & \mathbf{B}_{k}^{j,1}=\left\{ \left(\left(\omega,1\right),y^{1:\ell+1}\right)\right\} \\
\left(1-p_{1}^{S}\left(x^{\ell}\right)\right)\delta_{x^{1:\ell}}\left(y^{1:\ell}\right) & \mathbf{B}_{k}^{j,1}=\left\{ \left(\left(\omega,0\right),y^{1:\ell}\right)\right\} \\
0 & \mathrm{otherwise}
\end{cases}\label{eq:g_1_prediction}
\end{align}
 and, for $m>1$,
\begin{align}
 & g_{m}\left(\mathbf{B}_{k}^{j,m}\left|t_{k-1},\left\{ \left(\omega,x^{1:\ell}\right)\right\} \right.\right)\nonumber \\
 & =\begin{cases}
p_{m}^{S}\left(x^{\ell}\right)g_{m}\left(y|x^{\ell}\right) & \mathbf{B}_{k}^{j,m}=\left\{ \left(\left(\omega,m\right),y\right)\right\} \\
1-p_{m}^{S}\left(x^{\ell}\right) & \mathbf{B}_{k}^{j,m}=\emptyset\\
0 & \mathrm{otherwise}.
\end{cases}\label{eq:g_m_prediction}
\end{align}
The main branch can either die (we append 0 to $\omega$) or survive
(we append 1 to $\omega$ and the new state $y^{\ell+1}$ to previous
branch state). Potential spawning branches may spawn or not. If they
spawn they have a single state and we append its spawning mode $m$
to $\omega$. 

Finally, the birth model for tree trajectories at time step $k$ is
a PPP with intensity
\begin{align}
\lambda_{k}^{B}\left(t,\mathbf{B}\right) & =\begin{cases}
\delta_{k}[t]\lambda^{B}\left(x\right) & \mathbf{B}=\left\{ \left(1,x\right)\right\} \\
0 & \mathrm{otherwise}.
\end{cases}\label{eq:PPP_new_tree_trajectories}
\end{align}
It should be noted that all new born trajectories have a single branch
with genealogy variable $1$, as they only consist of the root node
of the tree.

\subsection{Measurement model for tree trajectories}

We can write the measurement model in Section \ref{subsec:Models_targets}
in terms of tree trajectories. To this end, we only need to write
the density of the target generated measurements as a function of
tree trajectories, the clutter model remains a PPP with intensity
$\lambda^{C}\left(\cdot\right)$. Each tree trajectory $X\in\mathbf{X}_{k}$
generates a multi-Bernoulli set of measurements with density
\begin{align}
f\left(\mathbf{z}|X\right) & =\sum_{\mathbf{z}_{1}\uplus...\uplus\mathbf{z}_{\left|\tau_{k}\left(X\right)\right|}=\mathbf{z}}\prod_{i=1}^{\left|\tau_{k}\left(X\right)\right|}f\left(\mathbf{z}_{i}|x_{i}\right)\label{eq:measurement_model_trajectory}
\end{align}
where $\tau_{k}\left(X\right)=\left\{ x_{1},...,x_{\tau_{k}\left(X\right)}\right\} $
is the set of targets in $X$ at time step $k$, $f\left(\cdot|x\right)$
is the Bernoulli density of the measurement set generated by target
$x$, 
\begin{align*}
f\left(\mathbf{z}|x\right) & =\begin{cases}
1-p^{D}\left(x\right) & \mathbf{z}=\emptyset\\
p^{D}\left(x\right)l\left(z|x\right) & \mathbf{z}=\left\{ z\right\} \\
0 & \left|\mathbf{z}\right|>1.
\end{cases}
\end{align*}

\section{Tree Poisson multi-Bernoulli mixture density\label{sec:TrPMBM_filter}}

In this section we explain the structure of the TrPMBM posterior.
In Section \ref{subsec:Exact-PMBM-posterior}, we explain the exact
PMBM posterior for sets of all tree trajectories, with data associations
at a tree level. In Section \ref{subsec:Approximate-PMBM-posterior},
we explain the proposed PMBM posterior with multi-Bernoulli branches
in each Bernoulli tree and data associations at a branch level. Section
\ref{subsec:Single-branch-densities} explains the form of the single
branch densities in the proposed PMBM filter.

\subsection{Exact PMBM posterior for sets of tree trajectories\label{subsec:Exact-PMBM-posterior}}

Tree trajectories are born independently following a PPP (\ref{eq:PPP_new_tree_trajectories}),
and each tree trajectory moves independently of the rest following
a Markovian process with single tree transition density (\ref{eq:transition_density_tree_all})
and probability of survival $p^{S}(X)=1$. The measurement model has
a multi-Bernoulli form (\ref{eq:measurement_model_trajectory}) with
PPP clutter. These models correspond to a standard multi-object dynamic
model \cite{Mahler_book14} and a measurement model of the type in
\cite{Angel21}, being the underlying single object space the tree
trajectory space. 

Therefore, the posterior and predicted densities are PMBMs and are
obtained via the recursion in \cite{Angel21} with the single tree
trajectory space and integrals (see Section \ref{sec:Problem-formulation})
instead of the single target ones. That is, the density of the set
of all tree trajectories at time step $k$ given the sequence of measurements
up to time step $k'=\left\{ k,k-1\right\} $ is a PMBM of the form
\cite{Angel21}

\begin{align}
f_{k|k'}\left(\mathbf{X}_{k}\right) & =\sum_{\mathbf{Y}\uplus\mathbf{W}=\mathbf{X}_{k}}f_{k|k'}^{\mathrm{p}}\left(\mathbf{Y}\right)f_{k|k'}^{\mathrm{mbm}}\left(\mathbf{W}\right)\label{eq:PMBM}\\
f_{k|k'}^{\mathrm{p}}\left(\mathbf{X}_{k}\right) & =e^{-\int\lambda_{k|k'}\left(X\right)dX}\prod_{X\in\mathbf{X}_{k}}\lambda_{k|k'}\left(X\right)\label{eq:PMBM_Poisson}
\end{align}
where $f_{k|k'}^{\mathrm{p}}\left(\cdot\right)$ is the PPP density,
which has intensity $\lambda_{k|k'}\left(\cdot\right)$ and represents
undetected tree trajectories, and $f_{k|k'}^{\mathrm{mbm}}\left(\cdot\right)$
is an MBM representing potential tree trajectories detected at some
point up to time step $k'$. The summation in (\ref{eq:PMBM}) is
taken over all mutually disjoint (and possibly empty) sets $\mathbf{Y}$
and $\mathbf{W}$ whose union is $\mathbf{X}_{k}$. The MBM density
is
\begin{align}
f_{k|k'}^{\mathrm{mbm}}\left(\mathbf{X}_{k}\right) & =\sum_{a\in\mathcal{A}_{k|k'}}w_{k|k'}^{a}\sum_{\uplus_{i=1}^{n_{k|k'}}\mathbf{X}_{k}^{i}=\mathbf{X}_{k}}\prod_{i=1}^{n_{k|k'}}f_{k|k'}^{i,a^{i}}\left(\mathbf{X}_{k}^{i}\right)\label{eq:MBM_exact}
\end{align}
where $n_{k|k'}$ is the number of potentially detected trees, $\mathcal{A}_{k|k'}$
is the set of global tree hypotheses, $w_{k|k'}^{a}$ is the weight
of global tree hypothesis $a=\left(a^{1},...,a^{n_{k|k'}}\right)$,
which contains indices to local tree hypotheses for each potentially
detected tree and $f_{k|k'}^{i,a^{i}}\left(\cdot\right)$ is the Bernoulli
density of the tree trajectory with local tree hypothesis $a^{i}.$
At each time step in each local hypothesis, we can associate more
than one measurement to each Bernoulli tree, as each of its alive
branches can generate one measurement, see (\ref{eq:measurement_model_trajectory}).
A global tree hypothesis $a$ indicates the measurements that are
associated to each Bernoulli tree at each time step. All details regarding
(\ref{eq:PMBM})-(\ref{eq:MBM_exact}) can be found in \cite[Sec. II.B]{Angel21}. 

\subsection{Approximate PMBM posterior for sets of tree trajectories\label{subsec:Approximate-PMBM-posterior}}

A drawback of the above PMBM recursion is that it is difficult to
handle the resulting Bernoulli tree densities because of the dependencies
between branches that arise in the prediction step, and the global
hypotheses defined at a tree level, i.e., associating measurements
to Bernoulli trees instead of branches. To obtain a computationally
efficient recursion (to be explained in Section \ref{sec:TrPMBM-filter-recursion}),
with global hypotheses defined at a branch level, we introduce the
following approximations:
\begin{itemize}
\item A1 The PPP represents alive branches without spawning.
\item A2 Each Bernoulli tree trajectory has a deterministic start time.
\item A3 After each prediction step, the set of branches in each Bernoulli
tree trajectory is multi-Bernoulli.
\end{itemize}
As spawning is a low probability event, we use A1 to simplify implementation
by discarding spawning events of potential targets we have never observed.
The absence of spawning transforms the trees into single branch trees.
A single branch tree can at most generate one measurement, and therefore,
A1 implies that each measurement generates a new Bernoulli, similar
to the PMBM recursion for point targets/trajectories \cite{Williams15b,Granstrom18}.
A newly created Bernoulli single branch tree may have multiple possible
start times, which give rise to a mixture in each Bernoulli single
tree density \cite[Eq. (28)]{Granstrom18}. After the Bernoulli tree
initialisation, all the mixture components in each Bernoulli tree
are predicted and updated with the same equations. To simplify the
filter implementation, we use A2 and take the most likely start time
when a new Bernoulli is created. Approximation A2 is also used in
the trajectory filters in \cite{Angel20_e}.

Approximations A1 and A2 imply that Bernoulli tree densities have
a single branch with a unique genealogy at the time step they are
created. In addition, we know the genealogies of these potential branches
for all Bernoulli trees. Approximation A3 is performed to simplify
the filtering recursion by discarding the dependencies among branches
within the same Bernoulli tree. It is performed at each prediction
step using the KLD minimisation that will be explained in Section
\ref{subsec:Prediction-via-KLD}. 

\subsubsection{Global branch hypotheses}

In (\ref{eq:MBM_exact}), global hypotheses are defined at a tree
level, in which, at each step, we can associate more than one measurement
to each Bernoulli tree. Using A1-A3, we can also define global hypotheses
at a branch level (global branch hypotheses), in which we can associate
at most one measurement to each potential branch in a Bernoulli tree.
The number of potential branches in the $i$-th tree is denoted by
$n_{k|k'}^{i}$ (see (\ref{eq:Maximum_number_branches})). Then, the
total number of potential branches is $N_{k|k'}=\sum_{i=1}^{n_{k|k'}}n_{k|k'}^{i}$.
A local hypothesis for tree $i$ and branch $j$ is denoted by an
index $\alpha^{i,j}\in\left\{ 1,...,h_{k|k'}^{i,j}\right\} $, where
$h_{k|k'}^{i,j}$ is the number of local hypotheses. 

A global branch hypothesis $\alpha$ is then a sequence of length
$N_{k|k'}$ containing the local hypothesis for each branch such that
$\alpha=\left(\alpha^{i,j}\right)$ with $i=1,..,n_{k|k'},j=1,..,n_{k|k'}^{i}$.
The set of global branch hypotheses is $\mathcal{D}_{k|k'}$ and is
defined as follows. We refer to measurement $z_{k}^{j}$ using the
pair $\left(k,j\right)$ and the set of all such pairs $\left(k,j\right)$
up to (and including) time step $k$ is $\mathcal{M}_{k}$. Then,
the set of measurement indices up to time step $k$ that correspond
to local hypothesis $\alpha^{i,j}$ is $\mathcal{M}_{k}^{i,j,\alpha^{i,j}}\subseteq\mathcal{M}_{k}$,
with at most one measurement per time step. The set $\mathcal{M}_{k}^{i,j,\alpha^{i,j}}$
is built recursively as will be explained in Section \ref{subsec:PMBM_Update}.
In a global hypothesis $\alpha\in\mathcal{D}_{k|k'},$ all measurements
must be associated to a local hypothesis, there can only be at most
one measurement associated to a local hypothesis per time step and,
there cannot be more than one local hypothesis associated with the
same measurement.

\subsubsection{Approximate MBM with global branch hypotheses}

Under A1-A3 and using global branch hypotheses, the MBM density (\ref{eq:MBM_exact})
can be written as
\begin{align}
f_{k|k'}^{\mathrm{mbm}}\left(\mathbf{X}_{k}\right) & =\sum_{\alpha\in\mathcal{D}_{k|k'}}w_{k|k'}^{\alpha}\sum_{\uplus_{i=1}^{n_{k|k'}}\mathbf{X}_{k}^{i}=\mathbf{X}_{k}}\prod_{i=1}^{n_{k|k'}}f_{k|k'}^{i,\alpha}\left(\mathbf{X}_{k}^{i}\right)\label{eq:MBM_approx}\\
w_{k|k'}^{\alpha} & \propto\prod_{i=1}^{n_{k|k}}\prod_{j=1}^{n_{k|k'}^{i}}w_{k|k'}^{i,j,\alpha}
\end{align}
where $f_{k|k'}^{i,\alpha}\left(\cdot\right)$ is the Bernoulli density
of the $i$-th tree for a global branch hypothesis $\alpha$, and
$w_{k|k'}^{i,j,\alpha}$ is the weight of the $j$-th branch in the
$i$-th tree in global branch hypothesis $\alpha$. 

Due to A1-A3, $f_{k|k'}^{i,\alpha}\left(\cdot\right)$ has a deterministic
start time $\overline{t}^{i}$ and its branches are multi-Bernoulli,
which means that $f_{k|k'}^{i,\alpha}\left(\cdot\right)$ is characterised
by $\overline{t}^{i}$ and $\left\{ \left(r_{k|k'}^{i,j,\alpha},p_{k|k'}^{i,j,\alpha}\left(\cdot\right)\right)\right\} _{j=1}^{n_{k|k'}^{i}}$
where $r_{k|k'}^{i,j,\alpha}$ and $p_{k|k'}^{i,j,\alpha}\left(\cdot\right)$
are the probability of existence and single branch density of the
$j$-th branch of tree $i$ under global hypothesis $\alpha$, respectively.
The set of branches in a tree thus follows a multi-Bernoulli distribution
\cite{Mahler_book14}, in which the event with no branches is mapped
to a no tree event, as a tree with no branches does not exist. With
this mapping, we can write $f_{k|k'}^{i,\alpha}\left(\cdot\right)$
as
\begin{align}
 & f_{k|k'}^{i,\alpha}\left(\mathbf{X}\right)\nonumber \\
 & =\begin{cases}
\delta_{\overline{t}^{i}}\left[t\right]\prod_{j=1}^{n_{k|k'}^{i}}p_{k|k'}^{i,j,\alpha}\left(\mathbf{B}^{j}\right) & \mathbf{X}=\left\{ \left(t,\mathbf{B}\right)\right\} ,|\mathbf{B}|>0\\
\prod_{j=1}^{n_{k|k'}^{i}}\left(1-r_{k|k'}^{i,j,\alpha}\right) & \mathbf{X}=\emptyset\\
0 & \mathrm{otherwise}
\end{cases}\label{eq:Bernoulli_tree_MB_branches}
\end{align}
\begin{align}
p_{k|k'}^{i,j,\alpha}\left(\mathbf{B}^{j}\right) & =\begin{cases}
r_{k|k'}^{i,j,\alpha}p_{k|k'}^{i,j,\alpha}\left(B\right) & \mathbf{B}^{j}=\left\{ B\right\} \\
1-r_{k|k'}^{i,j,\alpha} & \mathbf{B}^{j}=\emptyset\\
0 & \mathrm{otherwise}
\end{cases}\label{eq:Bernoulli_branch}
\end{align}
where $\overline{t}^{i}$ is the deterministic start time of the tree,
$p_{k|k'}^{i,j,\alpha}\left(\cdot\right)$ is the Bernoulli density
of the $j$-th branch of tree $i$ under global  hypothesis $\alpha$.
Given a set of branches $\mathbf{B}$, $\mathbf{B}^{j}$ is the set
for the $j$-th branch, which can have at most one element, see Section
\ref{subsec:Alternative-branch-integration}, and, as it is uniquely
determined, there is no sum over the subsets of $\mathbf{B}$ in (\ref{eq:Bernoulli_tree_MB_branches})
(see the contrast w.r.t. (\ref{eq:MBM_exact}), where there is a sum
over the subsets of $\mathbf{X}_{k}$).  It should be noted that
$r_{k|k'}^{i,j,\alpha}$, $p_{k|k'}^{i,j,\alpha}\left(\cdot\right)$
and $w_{k|k'}^{i,j,\alpha}$ only depend on element $\alpha^{i,j}$
of $\alpha$ but this is kept implicit for notational simplicity.

From (\ref{eq:Bernoulli_tree_MB_branches}), we note that the probability
of existence and single tree density of $f_{k|k'}^{i,\alpha}\left(\cdot\right)$
are
\begin{align}
r_{k|k'}^{i,\alpha} & =1-\prod_{j=1}^{n_{k|k'}^{i}}\left(1-r_{k|k'}^{i,j,\alpha}\right).\label{eq:Probality_existence_tree_MB_branches}\\
p_{k|k'}^{i,\alpha}\left(t,\mathbf{B}\right) & =\frac{\delta_{\overline{t}^{i}}\left[t\right]\prod_{j=1}^{n_{k|k'}^{i}}p_{k|k'}^{i,j,\alpha}\left(\mathbf{B}^{j}\right)}{r_{k|k'}^{i,\alpha}},\;|\mathbf{B}|>0.\label{eq:single_tree_density_MB_branches}
\end{align}
In addition, A3 relaxes the genealogy constraints of the branches,
see Section \ref{subsec:Integration}, as sampled trees from (\ref{eq:Bernoulli_tree_MB_branches})
do not necessarily meet the constraints.

\subsection{Single branch densities\label{subsec:Single-branch-densities}}

In general, the single branch density $p_{k|k'}^{i,j,\alpha}\left(\cdot\right)$
for the $i$-th Bernoulli tree, $j$-th branch and local hypothesis
$\alpha^{i,j}$, see (\ref{eq:Bernoulli_branch}), can be written
as
\begin{align}
 & p_{k|k'}^{i,j,\alpha}\left(\omega,x^{1:\ell}\right)\nonumber \\
 & =\sum_{\kappa=\overline{t}^{i,j}}^{k}\beta_{k|k'}^{i,j,\alpha}\left(\kappa\right)\delta_{\overline{\omega}_{k,\kappa}^{i,j}}\left[\omega\right]\delta_{\ell\left(\overline{\omega}_{k,\kappa}^{i,j}\right)}\left[\ell\right]p_{k|k'}^{i,j,\alpha}\left(x^{1:\ell};\kappa\right)\label{eq:single_branch_density}
\end{align}
where $\overline{t}^{i,j}$ is the start time of the branch, $\overline{\omega}_{k,\kappa}^{i,j}$
is the genealogy variable if the branch ends at time step $\kappa$,
$\ell\left(\overline{\omega}_{k,\kappa}^{i,j}\right)$ is the length
of the branch (see (\ref{eq:length_branch})), $\beta_{k|k'}^{i,j,\alpha}\left(\kappa\right)$
is the probability that this branch ends at time step $\kappa$, and
$p_{k|k'}^{i,j,\alpha}\left(\cdot;\kappa\right)$ is the density of
the states assuming the branch ends at time step $\kappa$. Note that
$\beta_{k|k'}^{i,j,\alpha}\left(\kappa\right)$ sums to one over $\kappa$. 

Due to A1, the intensity of the PPP only considers a single, alive
branch in each tree. Therefore,, $\lambda_{k|k'}\left(X\right)$ is
non-zero only for trees $X=\left(t,\left\{ \omega,x^{1:\ell}\right\} \right)$
with its main branch alive
\begin{align}
 & \lambda_{k|k'}\left(t,\left\{ \omega,x^{1:\ell}\right\} \right)\nonumber \\
 & =\sum_{\overline{t}=1}^{k}\delta_{\overline{t}}\left[t\right]\delta_{1_{k-\overline{t}+1}}\left[\omega\right]\delta_{k-\overline{t}+1}\left[\ell\right]\lambda_{k|k'}\left(x^{1:\ell};\overline{t}\right)\label{eq:intensity_PPP}
\end{align}
where $1_{\ell}$ is a sequence of length $\ell$ with all ones.

\section{TrPMBM filter recursion\label{sec:TrPMBM-filter-recursion}}

This section explains the filtering recursion for the proposed TrPMBM
filter, whose posterior density was explained in Section \ref{subsec:Approximate-PMBM-posterior}.
The prediction step for the PPP is explained in Section \ref{subsec:PPP-prediction}.
We derive the Bernoulli tree prediction with independent branches
via KLD minimisation in Section \ref{subsec:Prediction-via-KLD}.
The PMBM update is explained in Section \ref{subsec:PMBM_Update}.

\subsection{PPP prediction\label{subsec:PPP-prediction}}

Given $\lambda_{k-1|k-1}\left(\cdot\right)$ of the form (\ref{eq:intensity_PPP}),
we apply the transition density of the surviving branch, as we use
Approximation A1, and the PMBM prediction \cite{Williams15b,Angel18_b,Granstrom18}
to obtain
\begin{align}
 & \lambda_{k|k-1}\left(t,\left\{ \omega,x^{1:\ell}\right\} \right)=\lambda_{k}^{B}\left(t,\left\{ \omega,x^{1:\ell}\right\} \right)\nonumber \\
 & \qquad+\sum_{\overline{t}=1}^{k-1}\delta_{\overline{t}}\left[t\right]\delta_{1_{k-\overline{t}+1}}\left[\omega\right]\delta_{k-\overline{t}+1}\left[\ell\right]\lambda_{k|k-1}\left(x^{1:\ell};\overline{t}\right)\label{eq:PPP_prediction}
\end{align}
where $\lambda_{k}^{B}\left(\cdot\right)$ is given by (\ref{eq:PPP_new_tree_trajectories})
and
\begin{align*}
\lambda_{k|k-1}\left(x^{1:\ell+1};\overline{t}\right) & =g_{1}\left(x^{\ell+1}|x^{\ell}\right)p_{1}^{S}\left(x^{\ell}\right)\lambda_{k-1|k-1}\left(x^{1:\ell};\overline{t}\right).
\end{align*}

\subsection{Bernoulli tree prediction via KLD minimisation\label{subsec:Prediction-via-KLD}}

In the prediction step, the number of Bernoulli trees does not change
$n_{k|k-1}=n_{k-1|k-1}$ and we perform prediction for each tree,
independently of the other trees. After the prediction, the set of
branches in each tree is no longer multi-Bernoulli due to dependencies
introduced by the spawning process. Therefore, we perform a KLD minimisation
in each Bernoulli tree to approximate the set of branches as multi-Bernoulli,
see A3. 

Specifically, given $f_{k-1|k-1}^{i,\alpha}\left(\cdot\right)$ of
the form (\ref{eq:Bernoulli_tree_MB_branches}), the true predicted
Bernoulli $\widetilde{f}_{k|k-1}^{i,\alpha}\left(\cdot\right)$ is
calculated with the transition density (\ref{eq:transition_density_tree_all})
and probability of survival, $p^{S}(X)=1$, to produce \cite{Williams15b}

\begin{align}
\widetilde{f}_{k|k-1}^{i,\alpha}\left(\mathbf{X}\right) & =\begin{cases}
\widetilde{r}_{k|k-1}^{i,\alpha}\delta_{\overline{t}^{i}}\left[t\right]\widetilde{p}_{k|k-1}^{i,\alpha}\left(\mathbf{B}\right) & \mathbf{X}=\left\{ \left(t,\mathbf{B}\right)\right\} \\
1-\widetilde{r}_{k|k-1}^{i,\alpha} & \mathbf{X}=\emptyset
\end{cases}\label{eq:Bernoulli_tree-predicted_true}
\end{align}
where $\widetilde{r}_{k|k-1}^{i,\alpha}=r_{k-1|k-1}^{i,\alpha}$ and
\begin{align}
 & \widetilde{p}_{k|k-1}^{i,\alpha}\left(\mathbf{B}_{k}\right)\nonumber \\
 & =\int g\left(\mathbf{B}_{k}\left|X_{k-1}\right.\right)p_{k-1|k-1}^{i,\alpha}\left(X_{k-1}\right)dX_{k-1}\label{eq:Bernoulli_tree-predicted_true2}\\
 & =\frac{1}{\widetilde{r}_{k|k-1}^{i,\alpha}}\int_{|\mathbf{B}_{k-1}|>0}\prod_{j=1}^{n_{k-1|k-1}^{i}}\left[\prod_{m=1}^{\varrho}g_{m}\left(\mathbf{B}_{k}^{j,m}\left|\overline{t}^{i},\mathbf{B}_{k-1}^{j}\right.\right)\right.\nonumber \\
 & \quad\left.\times p_{k-1|k-1}^{i,j,\alpha}\left(\mathbf{B}_{k-1}^{j}\right)\right]\delta\mathbf{B}_{k-1}^{1:n_{k-1|k-1}^{i}}
\end{align}
where $p_{k-1|k-1}^{i,\alpha}\left(X_{k-1}\right)$ is given by (\ref{eq:single_tree_density_MB_branches}),
which requires $|\mathbf{B}_{k-1}|>0$.  We should note that $\widetilde{p}_{k|k-1}^{i,\alpha}\left(\emptyset\right)=0$,
as there is always at least one branch if the tree exists, and $g\left(\mathbf{B}_{k}\left|X_{k-1}\right.\right)$
corresponds to (\ref{eq:transition_density_tree_all}) without the
tree start time. 

Our aim is to find an approximation to $\widetilde{f}_{k|k-1}^{i,\alpha}\left(\cdot\right)$
with multi-Bernoulli branches of the form (\ref{eq:Bernoulli_tree_MB_branches}).
We obtain this approximation via KLD minimisation resulting in the
next propositions, which are proved in Appendix \ref{sec:AppendixB}
and \ref{sec:AppendixC}, respectively.
\begin{prop}
\label{prop:KLD_minimisation_prediction}Given the posterior tree
Bernoulli density $f_{k-1|k-1}^{i,\alpha}\left(\cdot\right)$ of the
form (\ref{eq:Bernoulli_tree_MB_branches}), the predicted tree Bernoulli
density $f_{k|k-1}^{i,\alpha}\left(\cdot\right)$ of the form (\ref{eq:Bernoulli_tree_MB_branches})
that minimises the KLD $D\left(\widetilde{f}_{k|k-1}^{i,\alpha}||f_{k|k-1}^{i,\alpha}\right)$,
where $\widetilde{f}_{k|k-1}^{i,\alpha}\left(\cdot\right)$ is the
true predicted density, see (\ref{eq:Bernoulli_tree-predicted_true}),
has a Bernoulli density for the $m$-th branch spawned from previous
branch $j$ given by

\begin{align}
p_{k|k-1}^{i,\left(j,m\right),\alpha}\left(\mathbf{B}_{k}^{j,m}\right) & =\int g_{m}\left(\mathbf{B}_{k}^{j,m}\left|\overline{t}^{i},\mathbf{B}_{k-1}^{j}\right.\right)\nonumber \\
 & \quad\times p_{k-1|k-1}^{i,j,\alpha}\left(\mathbf{B}_{k-1}^{j}\right)\delta\mathbf{B}_{k-1}^{j}.\label{eq:Prop_prediction_eq}
\end{align}
\end{prop}
Applying Proposition \ref{prop:KLD_minimisation_prediction} to the
single branch densities in (\ref{eq:single_branch_density}), we obtain
the following prediction step.
\begin{prop}
\label{prop:Bernoulli_tree_prediction}Given the Bernoulli tree density
$f_{k-1|k-1}^{i,\alpha}\left(\cdot\right)$ of the form (\ref{eq:Bernoulli_tree_MB_branches})
with single branch density (\ref{eq:single_branch_density}), the
predicted Bernoulli tree density $f_{k|k-1}^{i,\alpha}\left(\cdot\right)$
of the form (\ref{eq:Bernoulli_tree_MB_branches}) with single branch
density (\ref{eq:single_branch_density}) that minimises the KLD $D\left(\widetilde{f}_{k|k-1}^{i,\alpha}||f_{k|k-1}^{i,\alpha}\right)$
has $n_{k|k-1}^{i}=\varrho\cdot n_{k-1|k-1}^{i}$ potential branches
and the following parameters. For the surviving branch, $m=1$, the
parameters are: $r_{k|k-1}^{i,j,\alpha}=r_{k-1|k-1}^{i,j,\alpha}$
and
\begin{align}
\overline{\omega}_{k,\kappa}^{i,j} & =\begin{cases}
\left(\overline{\omega}_{k-1,k-1}^{i,j},1\right) & \kappa=k\\
\left(\overline{\omega}_{k,\kappa}^{i,j},0\right) & \kappa\in\{\overline{t}^{i,j},...,k-1\}
\end{cases}\label{eq:genealogy_i_j_kappa}
\end{align}
\begin{align}
 & p_{k|k-1}^{i,j,\alpha}\left(x^{1:\ell};\kappa\right)\nonumber \\
 & =\begin{cases}
p_{k-1|k-1}^{i,j,\alpha}\left(x^{1:\ell};\kappa\right) & \kappa\in\left\{ \overline{t}^{i,j},...,k-2\right\} \\
p_{k-1|k-1}^{i,j,\alpha}\left(x^{1:\ell};\kappa\right)\left(1-p_{1}^{S}\left(x^{\ell}\right)\right) & \kappa=k-1\\
p_{k-1|k-1}^{i,j,\alpha}\left(x^{1:\ell-1};k-1\right)\\
\quad\times p_{1}^{S}\left(x^{\ell-1}\right)g_{1}\left(x^{\ell}|x^{\ell-1}\right) & \kappa=k
\end{cases}\label{eq:prediction_density_m1_prop}
\end{align}
\begin{align}
\beta_{k|k-1}^{i,j,\alpha}(\kappa) & =\begin{cases}
\beta_{k-1|k-1}^{i,j,\alpha}(\kappa) & \kappa\in\left\{ \overline{t}^{i,j},...,k-2\right\} \\
\left(1-p^{S}\right)\beta_{k-1|k-1}^{i,j,\alpha}(\kappa) & \kappa=k-1\\
p^{S}\beta_{k-1|k-1}^{i,j,\alpha}(k-1) & \kappa=k
\end{cases}\label{eq:prediction_beta_m1_prop}
\end{align}
where
\begin{align}
p^{S}= & \int p_{k-1|k-1}^{i,j,\alpha}\left(x^{\ell};k-1\right)p_{1}^{S}\left(x^{\ell}\right)dx^{\ell}\label{eq:Average_ps}
\end{align}
and $p_{k-1|k-1}^{i,j,\alpha}\left(x^{\ell};k-1\right)$ denotes the
marginal density at the last state of the branch.

For the $m$-th branch spawning from previous branch $j$, the branch
index is $j^{*}=j+(m-1)n_{k-1|k-1}^{i}$ and the parameters are: $h^{i,j^{*}}=h^{i,j}$,
$\mathcal{M}_{k-1}^{i,j^{*},\alpha^{i,j^{*}}}=\emptyset$, $\beta_{k|k-1}^{i,j^{*},\alpha}(\kappa)=\delta_{k}[\kappa]$,
$\overline{t}^{i,j^{*}}=k$, and
\begin{align}
\overline{\omega}_{k,k}^{i,j^{*}} & =\left(\overline{\omega}_{k-1,k-1}^{i,j},m\right)\label{eq:genealogy_i_j_kappa_spawning}\\
p_{k|k-1}^{i,j^{*},\alpha}\left(y;k\right) & =\frac{\int g_{m}\left(y|x^{\ell}\right)p_{m}^{S}\left(x^{\ell}\right)p_{k-1|k-1}^{i,j,\alpha}\left(x^{\ell};k-1\right)dx^{\ell}}{\left\langle p_{m}^{S}\left(x^{\ell}\right),p_{k-1|k-1}^{i,j,\alpha}\left(x^{\ell};k-1\right)\right\rangle }\label{eq:prediction_density_m_higher1_prop}\\
r_{k|k-1}^{i,j^{*},\alpha} & =r_{k-1|k-1}^{i,j,\alpha}\left\langle p_{m}^{S}\left(x^{\ell}\right),p_{k-1|k-1}^{i,j,\alpha}\left(x^{\ell};k-1\right)\right\rangle \nonumber \\
 & \quad\times\beta_{k-1|k-1}^{i,j,\alpha}(k-1).\label{eq:prediction_existence_m_higher1_prop}
\end{align}
\end{prop}
Due to the KLD minimisation, we propagate each branch in each local
hypothesis independently of the rest. The surviving branch density
contains information on past states of the branch, and its predicted
density is similar to the standard trajectory case \cite{Angel20_e}.
The spawned branches only have one state, corresponding to the current
time step.

Proposition \ref{prop:KLD_minimisation_prediction} makes the TrPMBM
filter implementation considerably easier. If the spawning probability
is zero, the TrPMBM filter  computes the posterior density. Therefore,
the posterior and the TrPMBM approximation are expected to be similar
unless there is a high probability of a spawning event. Nevertheless,
if the spawning event happens at a time step $k_{1}$, the difference
between the posterior over the set of tree trajectories in a time
interval $\left[k_{1}+\Delta k,k\right]$ and its approximation should
vanish for a sufficiently large $\Delta k$, due to the forgetting
property of Markov systems \cite{Douc09}.

\subsection{Update\label{subsec:PMBM_Update}}

Given a global  hypothesis, the branches in a predicted PMBM density
with MBM of the form (\ref{eq:MBM_approx}), are multi-Bernoulli.
This implies that the updated density can be calculated similarly
to the PMBM update on point targets and trajectories \cite{Williams15b,Granstrom18},
by performing the data associations for the Bernoulli branches. The
resulting update is given in the following proposition.
\begin{prop}
\label{prop:PMBM_update}Under Approximations A1-A3, given a predicted
PMBM of the form (\ref{eq:PMBM}), (\ref{eq:PMBM_Poisson}) and (\ref{eq:MBM_approx}),
the updated density with measurement set $\mathbf{z}_{k}=\left\{ z_{k}^{1},...,z_{k}^{m_{k}}\right\} $
is a PMBM of the same form. The PPP intensity for undetected trees
is
\begin{align}
\lambda_{k|k}\left(t,\left\{ \omega,x^{1:\ell}\right\} \right) & =\left(1-p^{D}\left(x^{\ell}\right)\right)\lambda_{k|k-1}\left(t,\left\{ \omega,x^{1:\ell}\right\} \right)
\end{align}
where $\lambda_{k|k-1}\left(\cdot\right)$ is given by (\ref{eq:intensity_PPP}).

The number of updated Bernoulli tree components is $n_{k|k}=n_{k|k-1}+m_{k}$.
For each previous Bernoulli branch \textup{$j\in\left\{ 1,...,n_{k|k-1}^{i}\right\} $
}in a previous Bernoulli tree $i\in\left\{ 1,...,n_{k|k-1}\right\} $,
the update creates $\left(m_{k}+1\right)$ new local hypotheses corresponding
to a missed detection and an update with one of the received measurements,
which implies $h_{k|k}^{i,j}=h_{k|k-1}^{i,j}\left(m_{k}+1\right)$.
For missed detection hypotheses, $\alpha^{i,j}\in\left\{ 1,...,h_{k|k-1}^{i,j}\right\} ,$
the parameters are $\mathcal{M}_{k}^{i,j,\alpha^{i,j}}=\mathcal{M}_{k-1}^{i,j,\alpha^{i,j}}$

\begin{align}
w_{k|k}^{i,j,\alpha} & =w_{k|k-1}^{i,j,\alpha}\left(1-r_{k|k-1}^{i,j,\alpha}\beta_{k|k-1}^{i,j,\alpha}\left(k\right)p_{k}^{D,i,j,\alpha}\right)\\
r_{k|k}^{i,j,\alpha} & =\frac{r_{k|k-1}^{i,j,\alpha}\left(1-\beta_{k|k-1}^{i,j,\alpha}\left(k\right)p_{k}^{D,i,j,\alpha}\right)}{1-r_{k|k-1}^{i,j,\alpha}\beta_{k|k-1}^{i,j,\alpha}\left(k\right)p_{k}^{D,i,j,\alpha}}\\
\beta_{k|k}^{i,j,\alpha}\left(\kappa\right) & \propto\begin{cases}
\beta_{k|k-1}^{i,j,\alpha}\left(\kappa\right) & \overline{t}^{i,j}\leq\kappa<k\\
\left(1-p_{k}^{D,i,j,\alpha}\right)\beta_{k|k-1}^{i,j,\alpha}\left(k\right) & \kappa=k
\end{cases}
\end{align}
\begin{align}
p_{k|k}^{i,j,\alpha}\left(x^{1:\ell};\kappa\right) & =\begin{cases}
p_{k|k-1}^{i,j,\alpha}\left(x^{1:\ell};\kappa\right) & \overline{t}^{i,j}\leq\kappa<k\\
\frac{p_{D}\left(x^{\ell}\right)p_{k|k-1}^{i,j,\alpha}\left(x^{1:\ell};k\right)}{p_{k}^{D,i,j,\alpha}} & \kappa=k
\end{cases}
\end{align}
where
\begin{align}
p_{k}^{D,i,j,\alpha} & =\left\langle p^{D},p_{k|k-1}^{i,j,\alpha}\left(\cdot;k\right)\right\rangle \nonumber \\
 & =\int p^{D}\left(x^{\ell}\right)p_{k|k-1}^{i,j,\alpha}\left(x^{\ell};k\right)dx^{\ell}.\label{eq:expected_pd}
\end{align}

For a previous Bernoulli branch $j\in\left\{ 1,...,n_{k|k-1}^{i}\right\} $
in Bernoulli tree $i\in\left\{ 1,...,n_{k|k-1}\right\} $ with $\widetilde{\alpha}^{i,j}\in\left\{ 1,...,h_{k|k-1}^{i,j}\right\} $,
the new local hypothesis generated by measurement $z_{k}^{m}$ has
$\alpha^{i,j}=\widetilde{\alpha}^{i,j}+h_{k|k-1}^{i,j}m$, $r_{k|k}^{i,j,\alpha}=1$,
and 
\begin{align}
\mathcal{M}_{k}^{i,j,\alpha^{i,j}} & =\mathcal{M}_{k-1}^{i,j,\widetilde{\alpha}^{i,j}}\cup\left\{ \left(k,m\right)\right\} \\
w_{k|k}^{i,j,\alpha} & =w_{k|k-1}^{i,j,\widetilde{\alpha}}r_{k|k-1}^{i,j,\widetilde{\alpha}}\beta_{k|k-1}^{i,j,\widetilde{\alpha}}\left(k\right)l\left(z_{k}^{m}\right)\\
\beta_{k|k}^{i,j,\alpha}\left(\kappa\right) & =\begin{cases}
0 & \overline{t}^{i,j}\leq\kappa<k\\
1 & \kappa=k
\end{cases}\\
p_{k|k}^{i,j,\alpha}\left(x^{1:\ell};k\right) & \propto l\left(z_{k}^{m}|x^{\ell}\right)p^{D}\left(x^{\ell}\right)p_{k|k-1}^{i,j,\widetilde{\alpha}}\left(x^{1:\ell};k\right)\label{eq:Branch_update}\\
l\left(z_{k}^{m}\right) & =\left\langle l\left(z_{k}^{m}|\cdot\right)p^{D}\left(\cdot\right),p_{k|k-1}^{i,j,\widetilde{\alpha}}\left(\cdot;k\right)\right\rangle .\label{eq:Branch_update_normalising_constant}
\end{align}
Finally, the Bernoulli tree initiated by measurement $z_{k}^{m}$,
whose index is $i=n_{k|k-1}+m$, has one branch ($j=1$), $h_{k|k}^{i,1}=2$
local hypotheses, one with a non-existent Bernoulli
\begin{equation}
\mathcal{M}_{k}^{i,1,1}=\emptyset,\;w_{k|k}^{i,1,1}=1,\;r_{k|k}^{i,1,1}=0
\end{equation}
and the other with $\mathcal{M}_{k}^{i,1,2}=\left\{ z_{k}^{m}\right\} $,
and
\begin{align}
w_{k|k}^{i,1,2} & =\lambda^{C}\left(z_{k}^{m}\right)+\sum_{\overline{t}=1}^{k}\left\langle l\left(z_{k}^{m}|\cdot\right)p^{D},\lambda_{k|k-1}\left(\cdot;\overline{t}\right)\right\rangle \\
r_{k|k}^{i,1,2} & =\frac{\sum_{\overline{t}=1}^{k}\left\langle l\left(z_{k}^{m}|\cdot\right)p^{D},\lambda_{k|k-1}\left(\cdot;\overline{t}\right)\right\rangle }{w_{k|k}^{i,1,2}}\\
\overline{t}^{i} & =\underset{\overline{t}}{\arg\max}\left\langle l\left(z_{k}^{m}|\cdot\right)p^{D},\lambda_{k|k-1}\left(\cdot;\overline{t}\right)\right\rangle \label{eq:new_Bernoulli_tree_start_time}\\
p_{k|k}^{i,1,2}\left(\omega,x^{1:\ell}\right) & \propto l\left(z|x^{\ell}\right)p^{D}\left(x^{\ell}\right)\lambda_{k|k-1}\left(\overline{t}^{i},\left\{ \omega,x^{1:\ell}\right\} \right).\label{eq:new_Bernoulli_branch_density}
\end{align}
\end{prop}
As the predicted PPP is a mixture with different start times, see
(\ref{eq:intensity_PPP}), the newly created Bernoulli trees may have
multiple start times. To have a tree with deterministic start time
and simplify filter implementation, see A2 and (\ref{eq:single_tree_density_MB_branches}),
we take the most likely start time in (\ref{eq:new_Bernoulli_tree_start_time})
to obtain the density (\ref{eq:new_Bernoulli_branch_density}).

We would like to remark that, if targets are born according to a multi-Bernoulli
birth model instead of Poisson birth model, the posterior is a multi-Bernoulli
mixture (MBM), which is a PMBM with PPP intensity set to zero. The
MBM posterior can be computed with the same prediction and update
as in the PMBM filter (with PPP intensity equal to zero) adding the
Bernoulli components of new born targets in the prediction step \cite{Angel18_b,Xia19_b}.

\section{Gaussian TrPMBM filter recursion\label{sec:Gaussian-TrPMBM-filter}}

In this section, we explain the Gaussian implementation of the TrPMBM
filter for the linear/Gaussian model:
\begin{itemize}
\item $l\left(\cdot|x\right)=\mathcal{N}\left(\cdot;Hx,R\right)$ and $p^{D}\left(x\right)=p^{D}$.
\item $g_{m}\left(\cdot|x\right)=\mathcal{N}\left(\cdot;F_{m}x+d_{m},Q_{m}\right)$
and $p_{m}^{S}\left(x\right)=p_{m}^{S}$ for $m\in\{1,...,\varrho\}$.
\item $\lambda_{k}^{B}\left(x\right)=\sum_{q=1}^{n_{k}^{b}}w_{k}^{b,q}\mathcal{N}\left(x;\overline{x}_{k}^{b,q},P_{k}^{b,q}\right)$.
\end{itemize}
If we set $\varrho=1$, which implies there is no spawning, the Gaussian
TrPMBM filter implementation becomes the Gaussian trajectory PMBM
filter implementation for all trajectories in \cite{Angel20_e}, which
also uses Approximations A1 and A2. The single branch Gaussian densities
are explained in Section \ref{subsec:Single-branch-densities}. The
prediction step is explained in Sections \ref{subsec:PPP_prediction_Gauss_appendix}
and \ref{subsec:Gaussian-Bernoulli-tree}. The update is addressed
in Section \ref{subsec:Gaussian-implementation-update}. Practical
aspects and estimation are explained in Sections \ref{subsec:Practical_aspects}
and \ref{subsec:Estimation}.

\subsection{Single branch Gaussian densities\label{subsec:Single-branch-Gaussian-densities}}

We define a Gaussian density on a single branch space as
\begin{align}
\mathcal{N}\left(\omega,x^{1:\ell};\overline{\omega},\overline{x},P\right) & =\begin{cases}
\mathcal{N}\left(x^{1:\ell};\overline{x},P\right) & \omega=\overline{\omega},\ell=\ell\left(\overline{\omega}\right)\\
0 & \mathrm{otherwise}
\end{cases}\label{eq:Gaussian_branch}
\end{align}
where $\overline{\omega}$ is the genealogy, $\ell\left(\overline{\omega}\right)$
is the length of the branch (number of states in the branch), see
(\ref{eq:length_branch}), $\overline{x}$ is the $n_{x}\ell\left(\overline{\omega}\right)\times1$
mean vector and $P$ the $n_{x}\ell\left(\overline{\omega}\right)\times n_{x}\ell\left(\overline{\omega}\right)$
covariance.

Then, the single branch density (\ref{eq:single_branch_density})
on a branch $B=\left(\omega,x^{1:\ell}\right)$ can be written as
\begin{align}
p_{k|k'}^{i,j,\alpha}\left(B\right) & =\sum_{\kappa=\overline{t}^{i,j}}^{k}\beta_{k|k'}^{i,j,\alpha}\left(\kappa\right)\nonumber \\
 & \quad\times\mathcal{N}\left(B;\overline{\omega}_{k,\kappa}^{i,j},\overline{x}_{k|k'}^{i,j,\alpha}\left(\kappa\right),P_{k|k'}^{i,j,\alpha}\left(\kappa\right)\right)\label{eq:Gaussian_mixture_branch}
\end{align}
where $\overline{\omega}_{k,\kappa}^{i,j}$, $\overline{x}_{k|k'}^{i,j,\alpha}\left(\kappa\right)$
and $P_{k|k'}^{i,j,\alpha}\left(\kappa\right)$ are the genealogy,
mean and covariance of the $j$-th branch in the $i$-th tree, with
end time $\kappa$. The PPP intensity (\ref{eq:intensity_PPP}) is
\begin{align}
\lambda_{k|k'}\left(t,\left\{ B\right\} \right) & =\sum_{q=1}^{n_{k|k'}^{p}}w_{k|k'}^{p,q}\delta_{\overline{t}_{k|k'}^{p,q}}[t]\nonumber \\
 & \quad\times\mathcal{N}\left(B;1_{k-\overline{t}_{k|k'}^{p,q}+1},\overline{x}_{k|k'}^{p,q},P_{k|k'}^{p,q}\right)\label{eq:PPP_intensity_Gaussian}
\end{align}
where $n_{k|k'}^{p}$ is the number of PPP terms, $\overline{t}_{k|k'}^{p,q}$,
$w_{k|k'}^{p,q}$, $\overline{x}_{k|k'}^{p,q}$ and $P_{k|k'}^{p,q}$
are the start time, weight, mean and covariance of the $q$-th term.

It should be noted that the single branch Gaussian (\ref{eq:Gaussian_branch})
is analogous to the single trajectory Gaussian for filters based on
sets of trajectories \cite[Eq. (40)]{Angel20_e}, with the aditional
information of the genealogy variable $\overline{\omega}$. A similar
relation holds for the single branch density (\ref{eq:Gaussian_mixture_branch})
and PPP intensity (\ref{eq:PPP_intensity_Gaussian}), see Eq. (64)
and (43) in \cite{Angel20_e}. 

\subsection{Gaussian PPP prediction\label{subsec:PPP_prediction_Gauss_appendix}}

Given the birth intensity for targets, see Section \ref{sec:Gaussian-TrPMBM-filter},
the birth intensity for trees can be written as
\begin{align*}
\lambda_{k}^{B}\left(t,\left\{ B\right\} \right) & =\sum_{q=1}^{n_{k}^{b}}w_{k}^{b,q}\delta_{k}[t]\mathcal{N}\left(B;1,\overline{x}_{k}^{b,q},P_{k}^{b,q}\right)
\end{align*}
where the tree start time is $k$ and the genealogy variable for new
born branches is 1, see (\ref{eq:PPP_new_tree_trajectories}).

Let us consider the posterior PPP intensity at time step $k-1$ is
of the form (\ref{eq:PPP_intensity_Gaussian}). The predicted PPP
intensity is obtained using (\ref{eq:PPP_prediction}) and the Kalman
filter prediction step \cite{Sarkka_book13} to obtain
\begin{align}
 & \lambda_{k|k-1}\left(t,\left\{ B\right\} \right)\nonumber \\
 & =\lambda_{k}^{B}\left(t,\left\{ B\right\} \right)+p_{1}^{S}\sum_{q=1}^{n_{k-1|k-1}^{p}}w_{k-1|k-1}^{p,q}\nonumber \\
 & \times\delta_{\overline{t}_{k-1|k-1}^{p,q}}[t]\mathcal{N}\left(B;1_{k-\overline{t}_{k-1|k-1}^{p,q}+1},\overline{x}_{k|k-1}^{p,q},P_{k|k-1}^{p,q}\right)
\end{align}
where
\begin{align}
\overline{x}_{k|k-1}^{p,q} & =\left[\left(\overline{x}_{k-1|k-1}^{p,q}\right)^{T},\left(\overline{F}_{1}\overline{x}_{k-1|k-1}^{p,q}+d_{1}\right)^{T}\right]^{T}\label{eq:mean_pred_PPP_append}\\
P_{k|k-1}^{p,q} & =\left[\begin{array}{cc}
P_{k-1|k-1}^{p,q} & P_{k-1|k-1}^{p,q}\overline{F}_{1}^{T}\\
\overline{F}_{1}P_{k-1|k-1}^{p,q} & \overline{F}_{1}P_{k-1|k-1}^{p,q}\overline{F}_{1}^{T}+Q_{1}
\end{array}\right]\label{eq:cov_pred_PPP_append}\\
\overline{F}_{1} & =\left[0_{1,k-\overline{t}_{k-1|k-1}^{p,q}-1},1\right]\otimes F_{1}\label{eq:F_bar_1_append}
\end{align}
where $0_{n,m}$ is a zero matrix of size $n\times m$.

\subsection{Gaussian Bernoulli tree prediction\label{subsec:Gaussian-Bernoulli-tree}}

The prediction for each Bernoulli tree is given by Proposition \ref{prop:Bernoulli_tree_prediction}.
We consider that a single branch density in the PMBM posterior at
time $k-1$ is of the form (\ref{eq:Gaussian_mixture_branch}). Then,
for the linear Gaussian models, the predicted density for the main
branch is
\begin{align}
p_{k|k-1}^{i,j,\alpha}\left(B\right) & =\sum_{\kappa=\overline{t}^{i,j}}^{k}\beta_{k|k-1}^{i,j,\alpha}\left(\kappa\right)\nonumber \\
 & \times\mathcal{N}\left(B;\overline{\omega}_{k,\kappa}^{i,j},\overline{x}_{k|k-1}^{i,j,\alpha}\left(\kappa\right),P_{k|k-1}^{i,j,\alpha}\left(\kappa\right)\right)
\end{align}
where $\overline{\omega}_{k,\kappa}^{i,j}$ is given by (\ref{eq:genealogy_i_j_kappa}),
$\overline{x}_{k|k-1}^{i,j,\alpha}\left(\kappa\right)=\overline{x}_{k-1|k-1}^{i,j,\alpha}\left(\kappa\right)$
and $P_{k|k-1}^{i,j,\alpha}\left(\kappa\right)=P_{k-1|k-1}^{i,j,\alpha}\left(\kappa\right)$
for $\kappa<k$, and $\overline{x}_{k|k-1}^{i,j,\alpha}\left(k\right)$
and $P_{k|k-1}^{i,j,\alpha}\left(k\right)$ are obtained using (\ref{eq:mean_pred_PPP_append})
and (\ref{eq:cov_pred_PPP_append}) with $\overline{x}_{k-1|k-1}^{i,j,\alpha}\left(k-1\right)$
and $P_{k-1|k-1}^{i,j,\alpha}\left(k-1\right)$ instead of $\overline{x}_{k-1|k-1}^{p,q}$
and $P_{k-1|k-1}^{p,q}$. We also have that
\begin{align}
\beta_{k|k-1}^{i,j,\alpha}\left(\kappa\right) & =\begin{cases}
\beta_{k-1|k-1}^{i,j,\alpha}(\kappa) & \kappa\in\left\{ \overline{t}^{i,j},...,k-2\right\} \\
\left(1-p_{1}^{S}\right)\beta_{k-1|k-1}^{i,j,\alpha}(\kappa) & \kappa=k-1\\
p_{1}^{S}\beta_{k-1|k-1}^{i,j,\alpha}(k-1) & \kappa=k.
\end{cases}
\end{align}

For the $m$-th branch spawning from previous branch $j$, we obtain
$\overline{t}_{k}^{i,j^{*}}=k$, $\beta_{k|k-1}^{i,j^{*},\alpha}\left(k\right)=1$,
$\overline{\omega}_{k,\kappa}^{i,j^{*}}$ given by (\ref{eq:genealogy_i_j_kappa_spawning}),
and
\begin{align}
\overline{x}_{k|k-1}^{i,j^{*},\alpha}\left(k\right) & =\overline{F}_{m}\overline{x}_{k-1|k-1}^{i,j,\alpha}\left(k-1\right)+d_{m}\\
P_{k|k-1}^{i,j^{*},\alpha}\left(k\right) & =\overline{F}_{m}P_{k-1|k-1}^{i,j,\alpha}\left(k-1\right)F_{m}^{T}+Q_{m}\\
r_{k|k-1}^{i,j^{*},\alpha} & =r_{k-1|k-1}^{i,j,\alpha}p_{m}^{S}\beta_{k-1|k-1}^{i,j,\alpha}(k-1)
\end{align}
where $j^{*}$ is the index of the spawning branch, $\overline{F}_{m}$
is obtained analogously to (\ref{eq:F_bar_1_append}), but using $F_{m}$
and $\ell(\overline{\omega}_{k-1|k-1}^{i,j})$ instead of $F_{1}$
and $k-\overline{t}_{k-1|k-1}^{p,q}+1$.

\subsection{Gaussian implementation update\label{subsec:Gaussian-implementation-update}}

Due to the multi-Bernoulli branches in (\ref{eq:MBM_approx}) and
the form of the single branch densities (\ref{eq:Gaussian_mixture_branch})
and PPP intensity (\ref{eq:PPP_intensity_Gaussian}), the TrPMBM update
in Proposition \ref{prop:PMBM_update} with Gaussian models is equivalent
to the Gaussian TPMBM update used in \cite{Angel20_e}, so we do not
provide further details. We would like to remark that, in the Gaussian
implementation, the Bernoulli density of a new branch, see (\ref{eq:new_Bernoulli_tree_start_time})
and (\ref{eq:new_Bernoulli_branch_density}), uses the tree start
time, mean and covariance of the component with highest weight. The
analogous equations for trajectories are Eq. (60)-(63) in \cite{Angel20_e}.

\subsection{Practical aspects\label{subsec:Practical_aspects}}

To deal with covariance matrices of increasingly long branches, we
use the $L$-scan approximation \cite{Angel19_f,Angel20_e}. This
approximation sets the covariance matrices of the PPP and Bernoulli
branches as block diagonal
\begin{align}
P_{k|k} & \approx\mathrm{diag}\left(\tilde{P}_{k|k}^{\overline{t}},\tilde{P}_{k|k}^{\overline{t}+1},...,\tilde{P}_{k|k}^{k-L},\tilde{P}_{k|k}^{k-L+1:k}\right)\label{eq:L_scan_cov_approx}
\end{align}
where $\tilde{P}_{k|k}^{k-L+1:k}\in\mathbb{R}^{L\cdot n_{x}\times L\cdot n_{x}}$
is the joint covariance of the last $L$ time steps, $\tilde{P}_{k|k}^{k}\in\mathbb{R}^{n_{x}\times n_{x}}$
is the covariance of the branch state at time step $k$, and $\overline{t}$
is the branch start time. That is, branch states before the last $L$
time steps are approximated as independent, and are not updated with
new measurements. The rationale behind this approximation is that,
due to the Markovian nature, a current measurement does not have much
effect on past branch states sufficiently far in the past. The window
size $L$ acts as a trade-off between accuracy and computational complexity.
The higher $L$ becomes, the more accurate the approximation ($L=k$
being the exact filter), and the higher the computational burden.

As global and local hypotheses grow unboundedly with time, we apply
PMBM pruning techniques\footnote{More information on PMBM pruning can be found in the online multi-target
tracking course at https://www.youtube.com/watch?v=Q9fHowxNtN8.} \cite[Sec. V.D]{Granstrom20}. In PMBM pruning, global hypothesis
weights, PPP weights and probability of existences that are sufficiently
low are set to zero, and the corresponding global hypotheses, PPP
and Bernoulli components can be removed from the posterior. In the
proposed Gaussian TrPMBM implementation, we perform pruning analogously
to the pruning of the PMBM posterior on the set of all trajectories
in \cite[Sec. V.C]{Angel20_e}. In particular, in the update, we use
ellipsoidal gating \cite{Challa_book11} to discard unlikely measurement
to Bernoulli associations. In addition, for each previous global  hypothesis,
we use Murty's algorithm, in combination with the Hungarian algorithm,
to choose the $k=\left\lceil N_{h}\cdot w_{k|k-1}^{\alpha}\right\rceil $
new global  hypotheses with highest weights \cite{Murty68,Angel18_b}.
After the update, we discard global  hypotheses whose weight is below
a threshold $\Gamma_{mbm}$ and only keep the $N_{h}$ global  hypotheses
with highest weights. We also eliminate Bernoulli components whose
existence is below a threshold $\Gamma_{b}$ and the PPP components
whose weight is below $\Gamma_{p}$. If $\beta_{k|k}^{i,j,\alpha}\left(k\right)<\Gamma_{a}$,
we set $\beta_{k|k}^{i,j,\alpha}\left(k\right)=0$, which implies
that the branch is no longer updated or predicted, but it is still
a component of the posterior. 

Finally, we would like to point out that the Gaussian implementation,
presented for linear/Gaussian models, can be generalised to non-linear/non-Gaussian
models. To do so, one can approximate $p^{D}\left(\cdot\right)$ and
$p^{S}\left(\cdot\right)$ as constants with their value at the current
state mean and use a non-linear Kalman filter \cite{Bar-Shalom_book01,Sarkka_book13,Bell93,Angel15_c}
for the single branch prediction and update steps, see (\ref{eq:prediction_density_m1_prop}),
(\ref{eq:prediction_density_m_higher1_prop}), (\ref{eq:Branch_update})
and (\ref{eq:new_Bernoulli_branch_density}). It is also possible
to use importance Gaussian quadrature/sigma-points \cite{Elvira21}
to improve the normalising constant approximation in the update, see
(\ref{eq:Branch_update_normalising_constant}) and (\ref{eq:new_Bernoulli_tree_start_time}).
More information on these approaches in the context of multi-object
filtering can be found in \cite{Angel21_b}. Another possibility is
to implement the TrPMBM recursion via particle filtering \cite{Arulampalam02}.

\subsection{Estimation\label{subsec:Estimation}}

We can apply several estimators to a PMBM posterior (\ref{eq:PMBM})
to estimate the set $\mathbf{\hat{X}}_{k}$ of tree trajectories \cite[Sec. VI]{Angel18_b}.
We adapt Estimator 1 in \cite[Sec. VI]{Angel18_b} for tree trajectories.
We first take the global hypothesis $\alpha_{*}$ with highest weight,
and then, we obtain the most likely end time for the $j$-th branch
of the $i$-th tree
\begin{align}
\kappa_{*}^{i,j} & =\underset{\kappa}{\arg\max}\,\beta_{k|k}^{i,j,\alpha_{*}}\left(\kappa\right).
\end{align}

The estimated set of branches for the $i$-th tree is $\mathbf{\hat{B}}_{k}^{i}=\left\{ \left(\overline{\omega}_{k,\kappa_{*}^{i,j}}^{i,j},\overline{x}_{k|k}^{i,j,\alpha_{*}}\left(\kappa_{*}^{i,j}\right)\right):r_{k|k}^{i,j,\alpha_{*}}>\Gamma_{d}\right\} $,
which reports the genealogy variables and means with most likely length
of the Bernoulli branches whose existence probability is greater than
$\Gamma_{d}$. Then, the estimated set of trees is $\mathbf{\hat{X}}_{k}=\left\{ \left(\overline{t}^{i},\mathbf{\hat{B}}_{k}^{i}\right):\mathbf{\hat{B}}_{k}^{i}\neq\emptyset\right\} $,
which includes the trees that have at least one estimated branch.
A pseudocode of the Gaussian TrPMBM filter is provided in Algorithm
\ref{alg:TrPMBM_filter}.

\begin{algorithm}
\caption{\label{alg:TrPMBM_filter}Gaussian TrPMBM filter pseudocode}

{\fontsize{9}{9}\selectfont

\begin{algorithmic}     

\State- Set $\lambda_{0|0}\left(\cdot\right)=0$, $n_{0|0}=0$.

\For{$k=1$ to \textit{final time step} }

\State- \textit{Prediction}:

\State $\:$$\circ\,$Predict the PPP, see Sec. \ref{subsec:PPP_prediction_Gauss_appendix}.

\State $\:$$\circ\,$Predict the Bernoulli branches, see Sec. \ref{subsec:Gaussian-Bernoulli-tree}.

\State $\:$$\circ\,$Apply $L$-scan to all covariance matrices,
see (\ref{eq:L_scan_cov_approx}).

\State- \textit{Update}:

\State $\:$$\circ\,$Update single branch hypotheses with ellipsoidal
gating, see Prop. \ref{prop:PMBM_update}.

\State $\:$$\circ\,$Go through previous global hypotheses, obtaining
best ranked updated global hypotheses with Murty's algorithm, see
Sec. \ref{subsec:Practical_aspects}.

\State- Estimate the set of all tree trajectories, see Sec. \ref{subsec:Estimation}.

\State- Prune PPP components, global hypotheses, and Bernoulli branches,
see Sec. \ref{subsec:Practical_aspects}.

\EndFor

\end{algorithmic}

}
\end{algorithm}

\section{Simulations\label{sec:Simulations}}

In this section, we evaluate the proposed TrPMBM and TrMBM filters\footnote{Matlab code available at https://github.com/Agarciafernandez/MTT.}.
These filters have been implemented with the parameters (see Section
\ref{subsec:Practical_aspects}): $N_{h}=100$, $\Gamma_{mbm}=10^{-4}$,
$\Gamma_{p}=10^{-4}$, $\Gamma_{b}=10^{-4},$ $\Gamma_{a}=10^{-4}$,
gating threshold 15, $\Gamma_{d}=0.4,$ and $L\in\{1,5\}$. We have
also implemented the PMBM filter with spawning (S-PMBM) in \cite{Su21},
which uses recycling \cite{Williams12} to add the information on
spawned targets into the PPP. In addition, we have implemented two
filters that do not take into account spawning: the PMBM \cite{Williams15b,Angel18_b}
and the trajectory PMBM (TPMBM) filter \cite{Granstrom18,Angel20_e}.
The TPMBM provides information on the trajectory of each branch, but
not on the genealogy. The PMBM and S-PMBM filter do not provide genealogy
information either and estimate the trajectory of each branch sequentially
by linking estimates with the same auxiliary variable \cite{Angel20_e}.
All filters have been implemented with the same parameters as the
TrPMBM filter (where applicable). We have also tested the GLMB filter
with spawning, with a maximum of 1000 global hypotheses\footnote{Matlab code available at https://github.com/dsbryant/glmb-spawning.}
\cite{Bryant18}. All the units in this section are in the international
system. 

The single target state is $x=\left[p_{x},\dot{p}_{x},p_{y},\dot{p}_{y}\right]^{T}$,
which contains position and velocity in a two-dimensional plane. Targets
move with $p_{1}^{S}=0.99$ and the nearly constant velocity model
with $d_{1}$ a zero vector,
\begin{align*}
F_{1}=I_{2}\otimes\left(\begin{array}{cc}
1 & \tau\\
0 & 1
\end{array}\right),\quad Q_{1}=qI_{2}\otimes\left(\begin{array}{cc}
\tau^{3}/3 & \tau^{2}/2\\
\tau^{2}/2 & \tau
\end{array}\right),
\end{align*}
where $\tau=1$ and $q=0.01$. Targets can spawn with two modes ($\varrho=3$)
perpendicular to the current target direction, one in each direction.
A unit vector perpendicular to the target direction is $u_{\perp}=\left[-\dot{p}_{y},0,\dot{p}_{x},0\right]^{T}/\sqrt{\dot{p}_{x}^{2}+\dot{p}_{y}^{2}}$.
Then, $d_{2}=5\cdot u_{\perp}$, $d_{3}=-5\cdot u_{\perp}$, and
\[
F_{2}=\left(\begin{array}{cccc}
1 & 0 & 0 & -\tau\\
0 & 0 & 0 & -1\\
0 & \tau & 1 & 0\\
0 & 1 & 0 & 0
\end{array}\right),\quad F_{3}=\left(\begin{array}{cccc}
1 & 0 & 0 & \tau\\
0 & 0 & 0 & 1\\
0 & -\tau & 1 & 0\\
0 & -1 & 0 & 0
\end{array}\right)
\]
also, $Q_{1}=Q_{2}=Q_{3}$. The spawning probabilities of the two
modes are $p_{2}^{S}=p_{3}^{S}=0.01$. In the prediction step, we
approximate the value of $u_{\perp}$ at the predicted mean to perform
the prediction for the linear/Gaussian models, see Section \ref{subsec:Practical_aspects}.

We measure target position with 
\begin{align*}
H=\left(\begin{array}{cccc}
1 & 0 & 0 & 0\\
0 & 0 & 1 & 0
\end{array}\right),\quad R=\sigma^{2}I_{2},
\end{align*}
where $\sigma^{2}=4$, and $p_{D}=0.9$. The clutter intensity is
$\lambda^{C}\left(z\right)=\overline{\lambda}^{C}u_{A}\left(z\right)$
where $u_{A}\left(z\right)$ is a uniform density in $A=\left[0,600\right]\times\left[0,400\right]$
and $\overline{\lambda}^{C}=10$. The birth intensity is characterised
by $n_{k}^{b}=1$, $\overline{x}_{k}^{b,1}=\left[300,3,170,1\right]^{T}$
and $P_{k}^{b,1}=\mathrm{diag}\left(\left[160^{2},1,100^{2},1\right]\right)$,
and $w_{k}^{b,1}=0.08$. The TrMBM and GLMB filters use multi-Bernoulli
birth with one Bernoulli with existence probability $0.08$ and the
same mean and covariance matrix to match the PHD of the PPP birth
model \cite{Mahler_book14}.

\begin{figure}
\begin{centering}
\includegraphics[scale=0.6]{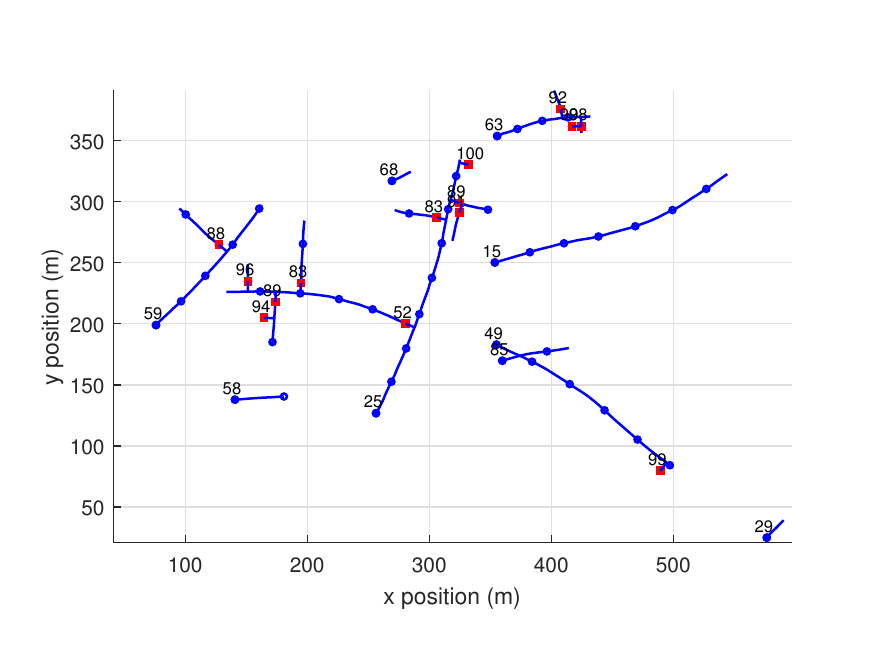}
\par\end{centering}
\caption{\label{fig:Scenario}Scenario of the simulations. Each tree trajectory
is shown in blue. Target positions every 10 time steps are marked
with a blue circle. A spawning event is marked with a red square.
The numbers next to a target birth or spawning event indicate the
corresponding time step. There are 9 trees and 23 branches.}
\end{figure}

The ground truth set of tree trajectories with $N_{s}=100$ time steps
has been obtained sampling from the dynamic model and is shown in
Figure \ref{fig:Scenario}. We evaluate the filters via Monte Carlo
simulation with $N_{mc}=100$ runs. To computes the estimation error,
we take the estimated set $\mathbf{\hat{X}}_{k}$ of trees and discard
tree and genealogy information to obtain an estimated set of trajectories
$\mathbf{\hat{T}}_{k}$, where we have one trajectory per each branch
\cite{Angel20_b}. We measure the error between $\mathbf{\hat{T}}_{k}$
and the ground truth set $\mathbf{T}_{k}$ of trajectories only considering
the positional elements. Error is computed with the linear programming
metric $d\left(\cdot,\cdot\right)$ for sets of trajectories in \cite{Angel20_d}
with parameters $p=2$, $c=10$ and $\gamma=1$, see Appendix \ref{sec:AppendixD}
for a review of its main characteristics. That is, the root mean square
(RMS) error at a given time step $k$, normalised by $k$ as in \cite{Angel20_e},
is
\begin{align}
d\left(k\right) & =\sqrt{\frac{1}{N_{mc}k}\sum_{i=1}^{N_{mc}}d^{2}\left(\mathbf{T}_{k},\mathbf{\hat{T}}_{k}^{i}\right)},\label{eq:error_time_k}
\end{align}
where $\mathbf{\hat{T}}_{k}^{i}$ is the estimated set of trajectories
at time $k$ in the $i$-th Monte Carlo run.

\begin{figure}
\begin{centering}
\includegraphics[scale=0.6]{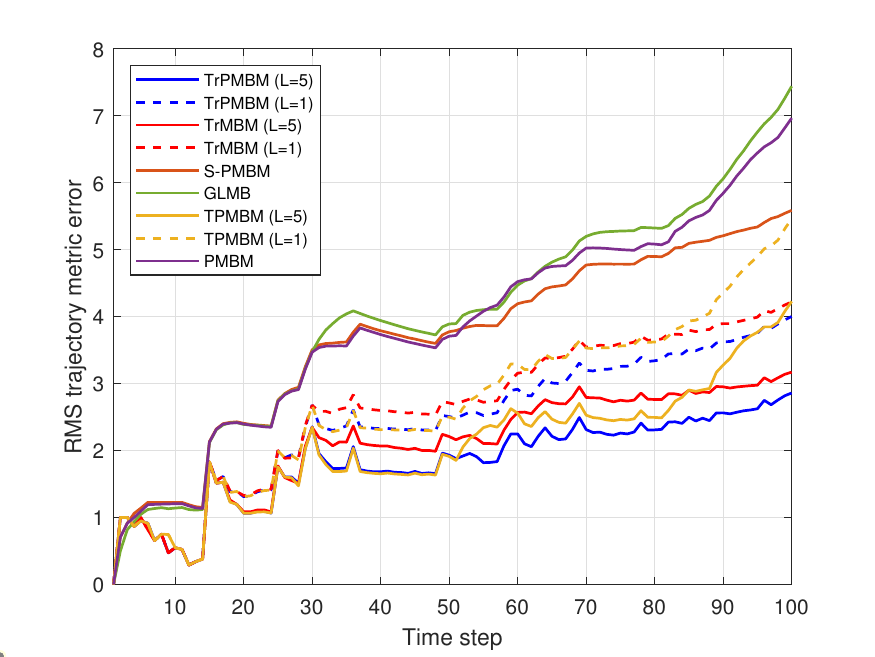}
\par\end{centering}
\caption{\label{fig:RMS-trajectory-metric}RMS trajectory metric error to estimate
all the branches at each time step. }

\end{figure}

We show the RMS trajectory metric error as a function of time in Figure
\ref{fig:RMS-trajectory-metric}. We can see that the best performing
filter is the TrPMBM ($L=5$) followed by the TrMBM ($L=5$) and the
TPMBM ($L=5$). This is reasonable as the TrPMBM filter approximates
the posterior density over the set of tree trajectories, which contains
full information on the trees. Behind these filters, we can find these
filters implemented with $L=1$, for which we expect lower performance
as they do not perform smoothing while filtering. 

The filters with lowest performance are the GLMB, the PMBM and the
S-PMBM. These filters estimate trajectories by linking target state
estimates with the same label or auxiliary variable. This approach
is inherently sub-optimal as it does not perform estimation of the
trajectories directly from a posterior density. The GLMB filter has
the lowest performance as it requires a higher number of global hypotheses
than a PMBM filter to represent the same information \cite[Sec. IV]{Angel18_b}\cite[App.D]{Angel20_e}.
The reason is that GLMB has global hypotheses with deterministic target
existence, while PMBM global hypotheses have probabilistic target
existence and undetected targets are efficiently represented in the
PMBM via a PPP. This results in an exponential increase in the number
of global hypotheses in the GLMB filter compared to the PMBM filter
\cite[Sec. IV]{Angel18_b}, and usually implies lower performance
and higher computational burden. The S-PMBM has better performance
than the PMBM filter as it accounts for target spawning.

\begin{figure}
\begin{centering}
\includegraphics[scale=0.3]{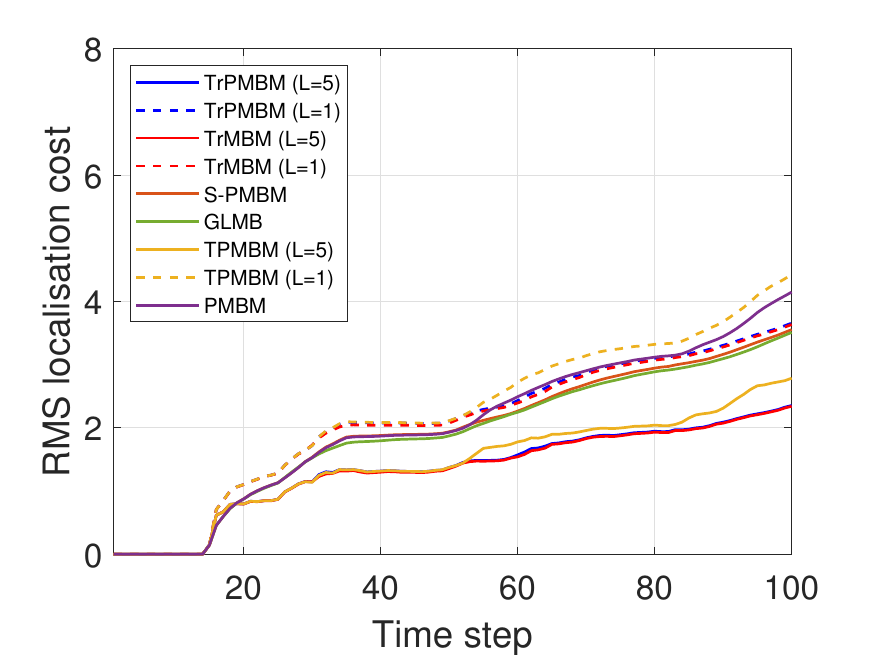}\includegraphics[scale=0.3]{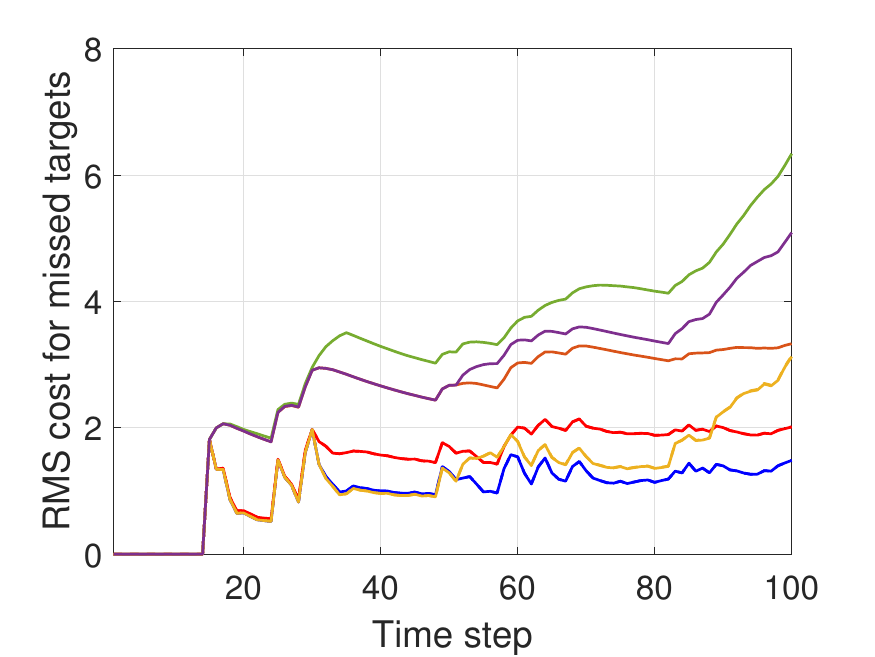}
\par\end{centering}
\begin{centering}
\includegraphics[scale=0.3]{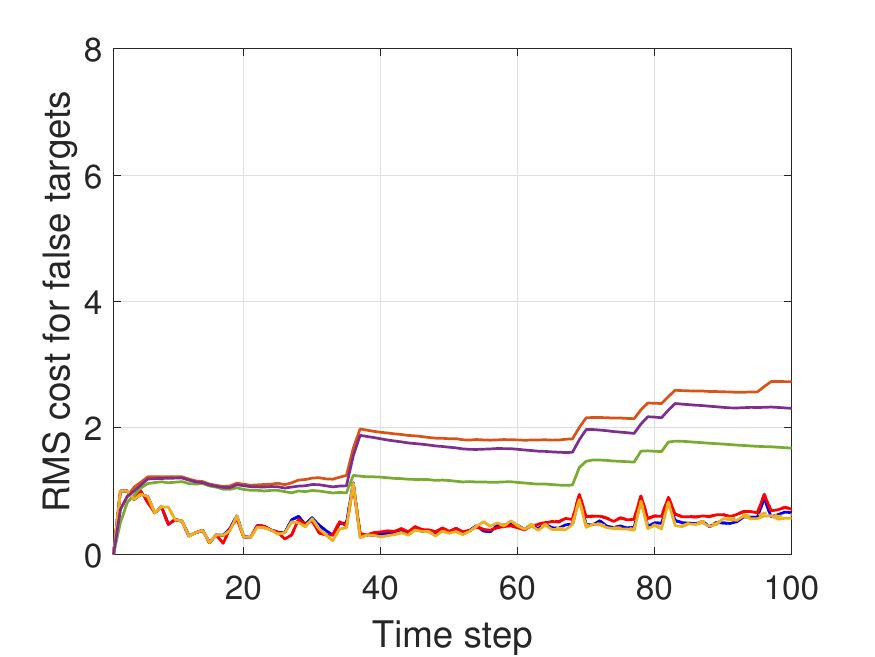}\includegraphics[scale=0.3]{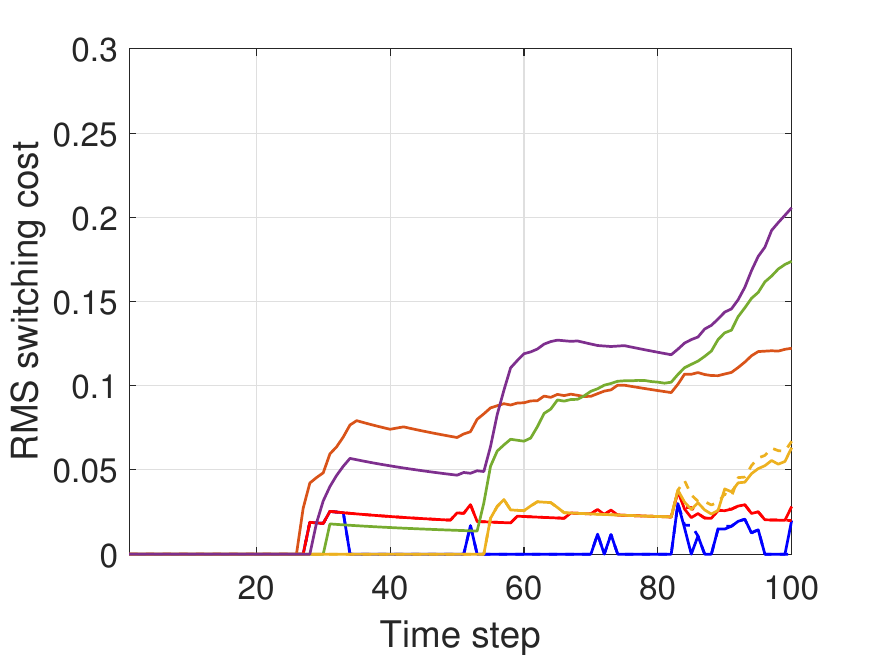}
\par\end{centering}
\caption{\label{fig:Trajectory_metric_decomposition}RMS trajectory metric
decomposition into localisation cost, missed target cost, false target
cost and track switching cost at each time step.}
\end{figure}

The decomposition of the RMS trajectory metric error into localisation,
missed target, false target and track switch costs is shown in Figure
\ref{fig:Trajectory_metric_decomposition}. The main difference between
TrPMBM and TrMBM is in the lower missed target cost of the TrPMBM.
This is a typical difference between PMBM and MBM filters \cite{Angel19_e},
due to the measurement-driven track initialisation in PMBM filtering.
Increasing $L$ lowers the localisation error of the trajectory and
tree trajectory filters. The reason is that with a higher value of
$L$, we can improve the estimation of past states of each trajectory,
improving localisation error. In this setting, increasing $L$ does
not affect the rest of the costs. TPMBM filters have quite similar
performance to TrPMBM up to time step 52, then performance worsens,
mainly due to localisation and missed target costs. This is expected
as the first target spawning is at time step 53, so trajectory and
tree trajectory filters work analogously up to this time step, and
then TrPMBM works better. GLMB, PMBM and S-PMBM have higher costs,
and the main difference in performance w.r.t. the other filters is
due to a higher number of false and missed targets, and they also
experience more track switches. For example, these sequential track
estimators may leave gaps in the estimated trajectories, even though
trajectories do have gaps according to the dynamic model. Tree or
trajectory filters do not leave gaps in estimated trajectories, as
the estimation is performed directly from the trajectory/tree posterior. 

The computational times of one Monte Carlo run of the filters on an
Intel core i5 laptop are provided in Table \ref{tab:Computational-times}.
The GLMB filter has the highest computational burden, due to its higher
number of global hypotheses, followed by TrPMBM and TrMBM. The TPMBM
filter is faster than TrPMBM and TrMBM filters, as TPMBM does not
take into account target spawning or tree information. The filters
with $L=1$ are faster than those with $L=5$, due to the faster single
branch updates. As expected, the PMBM filter is the fastest filter,
as it does not account for spawning and only keeps information on
the current set of targets.

\begin{table}
\caption{\label{tab:Computational-times}Average computational times in seconds
of the filters}

\begin{centering}
\begin{tabular}{c|cccccc}
\hline 
$L$ &
TrPMBM &
TrMBM &
TPMBM &
S-PMBM &
PMBM &
GLMB\tabularnewline
\hline 
$1$ &
9.8 &
6.0 &
2.6 &
\multirow{2}{*}{2.5} &
\multirow{2}{*}{1.5} &
\multirow{2}{*}{25.9}\tabularnewline
$5$ &
10.3 &
6.2 &
2.6 &  &  & \tabularnewline
\hline 
\end{tabular}
\par\end{centering}
\end{table}

\section{Conclusions\label{sec:Conclusions}}

In this paper, we have shown that we can obtain all information on
the trajectories of multiple spawning targets and their genealogies
by computing the posterior density on the set of all tree trajectories.
We have also shown that this posterior density is a PMBM, and we have
proposed a computationally efficient PMBM filter with independent
branches, derived via Kullback-Leibler divergence minimisation after
the prediction. 

It is possible to extend the TrPMBM recursion to extended targets,
and coexisting point and extended targets \cite{Granstrom20,Angel21}
by using the corresponding density for the set of measurements generated
by a single target. Further work includes the design of other types
of multi-target tracking algorithms for sets of tree trajectories.

\bibliographystyle{IEEEtran}
\bibliography{8C__Trabajo_laptop_Mis_articulos_Finished_Tree_PMBM_spawning_Accepted_Referencias}

\begin{thebibliography}{10}
\providecommand{\url}[1]{#1}
\csname url@samestyle\endcsname
\providecommand{\newblock}{\relax}
\providecommand{\bibinfo}[2]{#2}
\providecommand{\BIBentrySTDinterwordspacing}{\spaceskip=0pt\relax}
\providecommand{\BIBentryALTinterwordstretchfactor}{4}
\providecommand{\BIBentryALTinterwordspacing}{\spaceskip=\fontdimen2\font plus
\BIBentryALTinterwordstretchfactor\fontdimen3\font minus
  \fontdimen4\font\relax}
\providecommand{\BIBforeignlanguage}[2]{{%
\expandafter\ifx\csname l@#1\endcsname\relax
\typeout{** WARNING: IEEEtran.bst: No hyphenation pattern has been}%
\typeout{** loaded for the language `#1'. Using the pattern for}%
\typeout{** the default language instead.}%
\else
\language=\csname l@#1\endcsname
\fi
#2}}
\providecommand{\BIBdecl}{\relax}
\BIBdecl

\bibitem{Blackman_book99}
S.~Blackman and R.~Popoli, \emph{Design and Analysis of Modern Tracking
  Systems}.\hskip 1em plus 0.5em minus 0.4em\relax Artech House, 1999.

\bibitem{Gao19}
L.~Gao, G.~Battistelli, L.~Chisci, and P.~Wei, ``Distributed joint sensor
  registration and multitarget tracking via sensor network,'' \emph{IEEE
  Transactions on Aerospace and Electronic Systems}, vol.~56, no.~2, pp.
  1301--1317, 2020.

\bibitem{Chen19}
Q.~{Chen} and A.~{Tsukada}, ``Detection-by-tracking boosted online {3D}
  multi-object tracking,'' in \emph{IEEE Intelligent Vehicles Symposium}, 2019,
  pp. 295--301.

\bibitem{Chenouard14}
N.~Chenouard \emph{et~al.}, ``Objective comparison of particle tracking
  methods,'' \emph{Nature methods}, vol.~11, no.~3, pp. 281--290.

\bibitem{Magnusson15}
K.~E.~G. {Magnusson}, J.~{Jaldén}, P.~M. {Gilbert}, and H.~M. {Blau}, ``Global
  linking of cell tracks using the {V}iterbi algorithm,'' \emph{IEEE
  Transactions on Medical Imaging}, vol.~34, no.~4, pp. 911--929, 2015.

\bibitem{Mahler_book14}
R.~P.~S. Mahler, \emph{Advances in Statistical Multisource-Multitarget
  Information Fusion}.\hskip 1em plus 0.5em minus 0.4em\relax Artech House,
  2014.

\bibitem{Meyer18}
F.~{Meyer}, T.~{Kropfreiter}, J.~L. {Williams}, R.~{Lau}, F.~{Hlawatsch},
  P.~{Braca}, and M.~Z. {Win}, ``Message passing algorithms for scalable
  multitarget tracking,'' \emph{Proceedings of the IEEE}, vol. 106, no.~2, pp.
  221--259, Feb. 2018.

\bibitem{Angel20_b}
A.~F. García-Fernández, L.~Svensson, and M.~R. Morelande, ``Multiple target
  tracking based on sets of trajectories,'' \emph{IEEE Transactions on
  Aerospace and Electronic Systems}, vol.~56, no.~3, pp. 1685--1707, Jun. 2020.

\bibitem{Coraluppi14}
S.~Coraluppi and C.~A. Carthel, ``If a tree falls in the woods, it does make a
  sound: multiple-hypothesis tracking with undetected target births,''
  \emph{IEEE Transactions on Aerospace and Electronic Systems}, vol.~50, no.~3,
  pp. 2379--2388, July 2014.

\bibitem{Granstrom18}
K.~Granstr{ö}m, L.~Svensson, Y.~Xia, J.~L. Williams, and A.~F.
  García-Fernández, ``Poisson multi-{B}ernoulli mixture trackers: continuity
  through random finite sets of trajectories,'' in \emph{21st International
  Conference on Information Fusion}, 2018, pp. 973--981.

\bibitem{Williams15b}
J.~L. Williams, ``Marginal multi-{B}ernoulli filters: {RFS} derivation of
  {MHT}, {JIPDA} and association-based {MeMBer},'' \emph{IEEE Transactions on
  Aerospace and Electronic Systems}, vol.~51, no.~3, pp. 1664--1687, July 2015.

\bibitem{Angel18_b}
A.~F. García-Fernández, J.~L. Williams, K.~Granström, and L.~Svensson,
  ``Poisson multi-{B}ernoulli mixture filter: direct derivation and
  implementation,'' \emph{IEEE Transactions on Aerospace and Electronic
  Systems}, vol.~54, no.~4, pp. 1883--1901, Aug. 2018.

\bibitem{Xia19_b}
Y.~Xia, K.~Granstr{ö}m, L.~Svensson, A.~F. García-Fernández, and J.~L. Wlliams,
  ``Multi-scan implementation of the trajectory {P}oisson multi-{B}ernoulli
  mixture filter,'' \emph{Journal of Advances in Information Fusion}, vol.~14,
  no.~2, pp. 213--235, Dec. 2019.

\bibitem{Isaac08}
A.~Isaac, P.~Willett, and Y.~Bar-Shalom, ``Quickest detection and tracking of
  spawning targets using monopulse radar channel signals,'' \emph{IEEE
  Transactions on Signal Processing}, vol.~56, no.~3, pp. 1302--1308, 2008.

\bibitem{Dzyubachyk10}
O.~Dzyubachyk \emph{et~al.}, ``Advanced level-set-based cell tracking in
  time-lapse fluorescence microscopy,'' \emph{IEEE Transactions on Medical
  Imaging}, vol.~29, no.~3, pp. 852--867, 2010.

\bibitem{Flegel17}
S.~Flegel \emph{et~al.}, ``An analysis of the 2016 {H}itomi breakup event,''
  \emph{Earth Planets Space}, vol.~69, pp. 1--13.

\bibitem{Granstrom13b}
K.~Granström and U.~Orguner, ``On spawning and combination of extended/group
  targets modeled with random matrices,'' \emph{IEEE Transactions on Signal
  Processing}, vol.~61, no.~3, pp. 678--692, 2013.

\bibitem{Mahler03}
R.~P.~S. Mahler, ``Multitarget {B}ayes filtering via first-order multitarget
  moments,'' \emph{IEEE Transactions on Aerospace and Electronic Systems},
  vol.~39, no.~4, pp. 1152--1178, Oct. 2003.

\bibitem{Lundgren13}
M.~Lundgren, L.~Svensson, and L.~Hammarstrand, ``A {CPHD} filter for tracking
  with spawning models,'' \emph{IEEE Journal of Selected Topics in Signal
  Processing}, vol.~7, no.~3, pp. 496--507, June 2013.

\bibitem{Bryant17}
D.~S. {Bryant}, E.~D. {Delande}, S.~{Gehly}, J.~{Houssineau}, D.~E. {Clark},
  and B.~A. {Jones}, ``The {CPHD} filter with target spawning,'' \emph{IEEE
  Transactions on Signal Processing}, vol.~65, no.~5, pp. 13\,124--13\,138,
  2017.

\bibitem{Su21}
Z.~Su, H.~Ji, and Y.~Zhang, ``A {P}oisson multi-{B}ernoulli mixture filter with
  spawning based on {K}ullback-{L}eibler divergence minimization,''
  \emph{Chinese Journal of Aeronautics}, vol.~34, no.~11, pp. 154--168, 2021.

\bibitem{Bryant18}
D.~S. {Bryant}, B.-T. {Vo}, B.-N. {Vo}, and B.~A. {Jones}, ``A generalized
  labeled multi-{B}ernoulli filter with object spawning,'' \emph{IEEE
  Transactions on Signal Processing}, vol.~66, no.~23, pp. 6177--6189, 2018.

\bibitem{Nguyen21}
T.~T.~D. Nguyen, B.-N. Vo, B.-T. Vo, D.~Y. Kim, and Y.~S. Choi, ``Tracking
  cells and their lineages via labeled random finite sets,'' \emph{IEEE
  Transactions on Signal Processing}, vol.~69, pp. 5611--5625, 2021.

\bibitem{Xu21}
B.~Xu, M.~Lu, J.~Shi, J.~Cong, and B.~Nener, ``A joint tracking approach via
  ant colony evolution for quantitative cell cycle analysis,'' \emph{IEEE
  Journal of Biomedical and Health Informatics}, vol.~25, no.~6, pp.
  2338--2349, 2021.

\bibitem{Shi_book15}
Z.~Shi, \emph{Branching Random Walks}.\hskip 1em plus 0.5em minus 0.4em\relax
  Springer, 2015.

\bibitem{Popovic04}
L.~Popovic, ``Asymptotic genealogy of a critical branching process,'' \emph{The
  Annals of Applied Probability}, vol.~14, no.~4, pp. 2120--2148, Nov. 2004.

\bibitem{Angel21}
A.~F. García-Fernández, J.~L. Williams, L.~Svensson, and Y.~Xia, ``A {P}oisson
  multi-{B}ernoulli mixture filter for coexisting point and extended targets,''
  \emph{IEEE Transactions on Signal Processing}, vol.~69, pp. 2600--2610, 2021.

\bibitem{Mahler01}
R.~P.~S. Mahler, ``Detecting, tracking, and classifying group targets: a
  unified approach,'' in \emph{Proc. SPIE 4380}, 2001, pp. 217--228.

\bibitem{Swain10}
A.~Swain and D.~Clark, ``Extended object filtering using spatial independent
  cluster processes,'' in \emph{13th Conference on Information Fusion}, July
  2010, pp. 1--8.

\bibitem{Angel20_e}
A.~F. García-Fernández, L.~Svensson, J.~L. Williams, Y.~Xia, and
  K.~Granstr{ö}m, ``Trajectory {P}oisson multi-{B}ernoulli filters,''
  \emph{IEEE Transactions on Signal Processing}, vol.~68, pp. 4933--4945, 2020.

\bibitem{Douc09}
R.~Douc, G.~Fort, E.~Moulines, and P.~Priouret, ``Forgetting the initial
  distribution for hidden {M}arkov models,'' \emph{Stochastic Processes and
  their Applications}, vol. 119, no.~4, pp. 1235--1256, 2009.

\bibitem{Sarkka_book13}
S.~S{ä}rkk{ä}, \emph{Bayesian Filtering and Smoothing}.\hskip 1em plus 0.5em
  minus 0.4em\relax Cambridge University Press, 2013.

\bibitem{Angel19_f}
A.~F. García-Fernández and L.~Svensson, ``Trajectory {PHD} and {CPHD}
  filters,'' \emph{IEEE Transactions on Signal Processing}, vol.~67, no.~22,
  pp. 5702--5714, Nov 2019.

\bibitem{Granstrom20}
K.~Granstr{ö}m, M.~Fatemi, and L.~Svensson, ``Poisson multi-{B}ernoulli mixture
  conjugate prior for multiple extended target filtering,'' \emph{IEEE
  Transactions on Aerospace and Electronic Systems}, vol.~56, no.~1, pp.
  208--225, Feb. 2020.

\bibitem{Challa_book11}
S.~Challa, M.~R. Morelande, D.~Musicki, and R.~J. Evans, \emph{Fundamentals of
  Object Tracking}.\hskip 1em plus 0.5em minus 0.4em\relax Cambridge University
  Press, 2011.

\bibitem{Murty68}
K.~G. Murty, ``An algorithm for ranking all the assignments in order of
  increasing cost.'' \emph{Operations Research}, vol.~16, no.~3, pp. 682--687,
  1968.

\bibitem{Bar-Shalom_book01}
Y.~Bar-Shalom, T.~Kirubarajan, and X.~R. Li, \emph{Estimation with Applications
  to Tracking and Navigation}.\hskip 1em plus 0.5em minus 0.4em\relax John
  Wiley \& Sons, Inc., 2001.

\bibitem{Bell93}
B.~Bell and F.~Cathey, ``The iterated {K}alman filter update as a
  {G}auss-{N}ewton method,'' \emph{IEEE Transactions on Automatic Control},
  vol.~38, no.~2, pp. 294--297, Feb. 1993.

\bibitem{Angel15_c}
A.~F. García-Fernández, L.~Svensson, M.~R. Morelande, and S.~S{ä}rkk{ä},
  ``Posterior linearization filter: principles and implementation using sigma
  points,'' \emph{IEEE Transactions on Signal Processing}, vol.~63, no.~20, pp.
  5561--5573, Oct. 2015.

\bibitem{Elvira21}
V.~Elvira, L.~Martino, and P.~Closas, ``Importance {G}aussian quadrature,''
  \emph{IEEE Transactions on Signal Processing}, vol.~69, pp. 474--488, 2021.

\bibitem{Angel21_b}
A.~F. García-Fernández, J.~Ralph, P.~Horridge, and S.~Maskell, ``A {G}aussian
  filtering method for multi-target tracking with nonlinear/non-{G}aussian
  measurements,'' \emph{IEEE Transactions on Aerospace and Electronic Systems},
  vol.~57, no.~5, pp. 3539--3548, 2021.

\bibitem{Arulampalam02}
M.~Arulampalam, S.~Maskell, N.~Gordon, and T.~Clapp, ``A tutorial on particle
  filters for online nonlinear/non-{G}aussian {B}ayesian tracking,'' \emph{IEEE
  Transactions on Signal Processing}, vol.~50, no.~2, pp. 174--188, Feb. 2002.

\bibitem{Williams12}
J.~L. Williams, ``Hybrid {P}oisson and multi-{B}ernoulli filters,'' in
  \emph{15th International Conference on Information Fusion}, 2012, pp. 1103
  --1110.

\bibitem{Angel20_d}
A.~F. Garc{\'\i}a-Fern{\'a}ndez, A.~S. Rahmathullah, and L.~Svensson, ``A
  metric on the space of finite sets of trajectories for evaluation of
  multi-target tracking algorithms,'' \emph{IEEE Transactions on Signal
  Processing}, vol.~68, pp. 3917--3928, 2020.

\bibitem{Angel19_e}
A.~F. García-Fernández, Y.~Xia, K.~Granström, L.~Svensson, and J.~L. Williams,
  ``Gaussian implementation of the multi-{B}ernoulli mixture filter,'' in
  \emph{Proceedings of the 22nd International Conference on Information
  Fusion}, 2019.

\bibitem{Matheron_book75}
G.~Mathéron, \emph{Random Sets and Integral Geometry}.\hskip 1em plus 0.5em
  minus 0.4em\relax John Wiley \& Sons Inc, 1975.

\bibitem{Rahmathullah17}
A.~S. Rahmathullah, A.~F. García-Fernández, and L.~Svensson, ``Generalized
  optimal sub-pattern assignment metric,'' in \emph{20th International
  Conference on Information Fusion}, 2017, pp. 1--8.

\end{thebibliography}
\cleardoublepage{}

{\LARGE{}Supplementary material: ``Tracking multiple spawning targets
using Poisson multi-Bernoulli mixtures on sets of tree trajectories''}{\LARGE\par}

\appendices{}

\section{\label{sec:AppendixA}}

This appendix provides more details on trees and branches. 

\subsection{Length of a branch\label{subsec:Branch-length}}

The mathematical formula that defines the length of a branch with
genealogy $\omega=\left(\omega^{1},...,\omega^{\nu}\right)$ is
\begin{equation}
\ell\left(\omega\right)=e\left(\omega\right)-i\left(\omega\right)+1\label{eq:length_branch}
\end{equation}
where
\begin{align}
e\left(\omega\right)=\begin{cases}
\nu & \omega^{i'}\neq0,\forall i'\\
\min\left(i':\omega^{i'}=0\right)-1 & \mathrm{otherwise}
\end{cases}\\
i\left(\omega\right)=\begin{cases}
1 & \omega^{i'}\leq1,\forall i'\\
\max\left(i':\omega^{i'}>1\right) & \mathrm{otherwise}.
\end{cases}\label{eq:initial_generation_last_spawning}
\end{align}
We would like to remark that $i\left(\omega\right)$ is the generation
when the branch was spawned or born, and $e\left(\omega\right)$ is
the last generation when the branch is present.

\subsection{Genealogy constraints\label{subsec:Genealogy-constraints}}

The genealogy variables in a tree trajectory are subject to two constraints:
uniqueness and consistent offspring.  Let us consider the tree trajectory
\begin{align}
X & =\left(t,\left\{ \left(\omega_{1},x_{1}^{1:\ell_{1}}\right),...,\left(\omega_{n},x_{n}^{1:\ell_{n}}\right)\right\} \right)
\end{align}
with $\omega_{i}=\left(\omega_{i}^{1},...,\omega_{i}^{\nu}\right)$.
Given $\omega_{i}$, its unique identifier $\varpi_{i}$, see Definition
\ref{def:Branch-ID}, is
\begin{align}
\varpi_{i} & =\left(\omega_{i}^{1},...,\omega_{i}^{i\left(\omega_{i}\right)}\right)
\end{align}
where $i\left(\omega_{i}\right)$ is given by (\ref{eq:initial_generation_last_spawning}).
Uniqueness means that there cannot be more than one branch with a
unique identifier: $\nexists\,i,j$, $i\neq j$, such that $\varpi_{i}=\varpi_{j}$.
In addition, a tree must have a main branch, with unique identifier
1, which also implies $n>0$.

The property of consistent offspring is as follows. For each $i$
and $k$ such that $\omega_{i}^{k}>1$ (i.e., there is a spawning
event in branch $i$ at generation $k$), there is a $j\neq i$ such
that $\omega_{j}^{1:k-1}=\omega_{i}^{1:k-1}$ and $\omega_{j}^{k}\in\left\{ 0,1\right\} $.
That is, if there is branch $i$ with spawning at generation $k$,
there must be a parent branch $j$, which has the same genealogy up
to generation $k-1$, and then either it survives (with its main mode)
$\omega_{j}^{k}=1$ or it terminates $\omega_{j}^{k}=0$ at generation
$k$. 

\subsection{Tree trajectory space is LCHS\label{subsec:Tree-trajectory-LCHS}}

In this appendix, we explain why the space of single tree trajectories
is locally compact, Hausdorff and second-countable (LCHS) \cite{Mahler_book14,Angel20_b,Xia19_b}.
The space of a branch $\mathbb{B}_{(\nu)}=\uplus_{\omega\in I_{(\nu)}}\left\{ \omega\right\} \times\mathbb{R}^{\ell\left(\omega\right)\cdot n_{x}}$
is LCHS. The proof is similar to the proof that the single trajectory
space is LCHS \cite[App. A]{Angel20_b}. The space $\mathcal{F}\left(\mathbb{B}_{(k-t+1)}\right)$
is compact, Hausdorff and second-countable, which implies that it
is LCHS, see \cite[Thm. 1-2-1]{Matheron_book75}. Finally, as $\mathbb{T}_{(k)}$
is the disjoint union of LCHS spaces, we can also follow \cite[App. A]{Angel20_b}
to show that $\mathbb{T}_{(k)}$ is LCHS.

\subsection{Explicit single tree integral\label{subsec:App_explicit-single-tree-integral}}

By expanding the set integral inside the single tree integral in (\ref{eq:tree_trajectory_integral_general}),
we can write (\ref{eq:tree_trajectory_integral_general}) as \cite{Mahler_book14}
\begin{align}
 & \int\pi\left(X\right)dX\nonumber \\
 & =\sum_{t=1}^{k}\sum_{n=0}^{n_{max}\left(k-t+1\right)}\frac{1}{n!}\sum_{\omega_{1:n}\in I_{(k-t+1)}^{n}}\int\nonumber \\
 & \quad\pi\left(t,\left\{ \left(\omega_{1},x_{1}^{1:\ell_{1}}\right),...,\left(\omega_{n},x_{n}^{1:\ell_{n}}\right)\right\} \right)d\left(x_{1}^{1:\ell_{1}}...x_{n}^{1:\ell_{n}}\right)
\end{align}
where the branch length $\ell_{i}$ is determined by $\omega_{i}$,
i.e., $\ell_{i}=\ell\left(\omega_{i}\right)$, see Appendix \ref{subsec:Branch-length},
$n_{max}\left(\cdot\right)$ is given by (\ref{eq:Maximum_number_branches}),
and $I_{(\nu)}^{n}$ denotes the $n$-th Cartesian power of $I_{(\nu)}$.

\section{\label{sec:AppendixB}}

In this appendix we prove Proposition \ref{prop:KLD_minimisation_prediction}.
We use $\mathrm{z}$ to denote constants whose values do not affect
the minimisation. The products with indices $j$ and $m$ go through
$j\in\left\{ 1,...,n_{k-1|k-1}^{i}\right\} $ and $m\in\left\{ 1,...,\varrho\right\} $.
The KLD is \cite{Mahler_book14}
\begin{align}
 & D\left(\widetilde{f}_{k|k-1}^{i,\alpha}||f_{k|k-1}^{i,\alpha}\right)\nonumber \\
 & =\int\widetilde{f}_{k|k-1}^{i,\alpha}\left(\mathbf{X}_{k}\right)\log\frac{\widetilde{f}_{k|k-1}^{i,\alpha}\left(\mathbf{X}_{k}\right)}{f_{k|k-1}^{i,\alpha}\left(\mathbf{X}_{k}\right)}\delta\mathbf{X}_{k}\nonumber \\
 & =\mathrm{z}-\prod_{j}\left(1-r_{k-1|k-1}^{i,j,\alpha}\right)\log\prod_{j,m}\left(1-r_{k|k-1}^{i,\left(j,m\right),\alpha}\right)\nonumber \\
 & -\sum_{t}\int\widetilde{f}_{k|k-1}^{i,\alpha}\left(\left\{ \left(t,\mathbf{B}_{k}\right)\right\} \right)\log f_{k|k-1}^{i,\alpha}\left(\left\{ \left(t,\mathbf{B}_{k}\right)\right\} \right)\delta\mathbf{B}_{k}\nonumber \\
 & =\mathrm{z}-\prod_{j}\left(1-r_{k-1|k-1}^{i,j,\alpha}\right)\log\prod_{j,m}\left(1-r_{k|k-1}^{i,\left(j,m\right),\alpha}\right)\nonumber \\
 & -\widetilde{r}_{k|k-1}^{i,\alpha}\int_{|\mathbf{B}_{k}|>0}\widetilde{p}_{k|k-1}^{i,\alpha}\left(\mathbf{B}_{k}\right)\nonumber \\
 & \times\log\left[\prod_{j,m}p_{k|k-1}^{i,\left(j,m\right),\alpha}\left(\mathbf{B}_{k}^{j,m}\right)\right]\delta\mathbf{B}_{k}\nonumber \\
 & =\mathrm{z}-\prod_{j}\left(1-r_{k-1|k-1}^{i,j,\alpha}\right)\log\left[\prod_{j,m}\left(1-r_{k|k-1}^{i,\left(j,m\right),\alpha}\right)\right]\nonumber \\
 & -\int_{|\mathbf{B}_{k}|>0}\int_{|\mathbf{B}_{k-1}|>0}\prod_{j,m}g_{m}\left(\mathbf{B}_{k}^{j,m}\left|\overline{t}^{i},\mathbf{B}_{k-1}^{j}\right.\right)\nonumber \\
 & \times p_{k-1|k-1}^{i,j,\alpha}\left(\mathbf{B}_{k-1}^{j}\right)\delta\mathbf{B}_{k-1}^{1:n_{k-1|k-1}^{i}}\nonumber \\
 & \times\log\left[\prod_{j,m}p_{k|k-1}^{i,\left(j,m\right),\alpha}\left(\mathbf{B}_{k}^{j,m}\right)\right]\delta\mathbf{B}_{k}.\label{eq:KLD_append1}
\end{align}
According to the dynamic model and the posterior at time $k-1$, the
following equalities hold
\begin{align}
\widetilde{f}_{k|k-1}^{i,\alpha}\left(\emptyset\right) & =\prod_{j}\left(1-r_{k-1|k-1}^{i,j,\alpha}\right)\nonumber \\
 & =\prod_{j,m}g_{m}\left(\emptyset\left|\overline{t}^{i},\emptyset\right.\right)p_{k-1|k-1}^{i,j,\alpha}\left(\emptyset\right).
\end{align}
That is, the predicted tree is empty if there are no branches at time
step $k-1$, which implies no branches at time step $k$. Therefore,
we can remove the domain of integration $|\mathbf{B}_{k}|>0$ and
$|\mathbf{B}_{k-1}|>0$ by incorporating the previous term in (\ref{eq:KLD_append1})
into the integral to obtain
\begin{align*}
 & D\left(\widetilde{f}_{k|k-1}^{i,\alpha}||f_{k|k-1}^{i,\alpha}\right)\\
 & =\mathrm{z}-\int\int\prod_{j,m}g_{m}\left(\mathbf{B}_{k}^{j,m}\left|\overline{t}^{i},\mathbf{B}_{k-1}^{j}\right.\right)p_{k-1|k-1}^{i,j,\alpha}\left(\mathbf{B}_{k-1}^{j}\right)\\
 & \times\delta\mathbf{B}_{k-1}^{1:n_{k-1|k-1}^{i}}\sum_{j,m}\log\left[p_{k|k-1}^{i,\left(j,m\right),\alpha}\left(\mathbf{B}_{k}^{j,m}\right)\right]\delta\mathbf{B}_{k}\\
 & =\mathrm{z}-\sum_{j,m}\int\int g_{m}\left(\mathbf{B}_{k}^{j,m}\left|\overline{t}^{i},\mathbf{B}_{k-1}^{j}\right.\right)p_{k-1|k-1}^{i,j,\alpha}\left(\mathbf{B}_{k-1}^{j}\right)\\
 & \times\delta\mathbf{B}_{k-1}^{j}\log\left[p_{k|k-1}^{i,\left(j,m\right),\alpha}\left(\mathbf{B}_{k}^{\left(j,m\right)}\right)\right]\delta\mathbf{B}_{k}^{j,m}.
\end{align*}
By standard KLD minimisation we obtain (\ref{eq:Prop_prediction_eq}),
which proves Proposition \ref{prop:KLD_minimisation_prediction}.

\section{\label{sec:AppendixC}}

In this appendix we prove Proposition \ref{prop:Bernoulli_tree_prediction}.
We evaluate (\ref{eq:Prop_prediction_eq}) at $\mathbf{B}_{k}^{j,m}=\left\{ B\right\} $

\begin{align}
p_{k|k-1}^{i,\left(j,m\right),\alpha}\left(\left\{ B\right\} \right) & =\int g_{m}\left(\left\{ B\right\} \left|\overline{t}^{i},\left\{ B'\right\} \right.\right)p_{k-1|k-1}^{i,j,\alpha}\left(B'\right)dB'.\label{eq:Prediction_branch_append}
\end{align}
We proceed to analyse the cases $m=1$ and $m>1$.

\subsection{Case $m=1$}

The probability of existence of $p_{k|k-1}^{i,(j,1),\alpha}\left(\cdot\right)$
in (\ref{eq:Prediction_branch_append}) is
\begin{align}
r_{k|k-1}^{i,(j,1),\alpha} & =\int p_{k|k-1}^{i,(j,1),\alpha}\left(\left\{ B\right\} \right)dB\nonumber \\
 & =r_{k-1|k-1}^{i,j,\alpha}\label{eq:r_j_1_append}
\end{align}
where we have used that $g_{1}\left(\cdot|\cdot\right)$ in (\ref{eq:g_1_prediction})
does not change cardinality. The predicted single branch density is
\begin{alignat}{1}
p_{k|k-1}^{i,(j,1),\alpha}\left(B\right) & =\frac{p_{k|k-1}^{i,(j,1),\alpha}\left(\left\{ B\right\} \right)}{r_{k|k-1}^{i,(j,1),\alpha}}\\
 & =\int g_{1}\left(\left\{ B\right\} \left|t_{k-1},\left\{ B'\right\} \right.\right)p_{k-1|k-1}^{i,j,\alpha}\left(B'\right)dB'.\label{eq:integral_Appendix_C}
\end{alignat}
We calculate (\ref{eq:integral_Appendix_C}) using (\ref{eq:Bernoulli_branch_integral}),
(\ref{eq:g_1_prediction_dead}), (\ref{eq:g_1_prediction}), and (\ref{eq:single_branch_density})
to obtain
\begin{align}
 & p_{k|k-1}^{i,(j,1),\alpha}\left(B\right)\nonumber \\
 & =\sum_{\kappa=\overline{t}^{i,j}}^{k-1}\int g_{1}\left(\left\{ B\right\} \left|t_{k-1},\left\{ \left(\omega,y^{1:\ell}\right)\right\} \right.\right)\beta_{k-1|k-1}^{i,j,\alpha}\left(\kappa\right)\nonumber \\
 & \,\times\delta_{\overline{\omega}_{k-1,\kappa}^{i,j}}\left[\omega\right]\delta_{\ell(\overline{\omega}_{k-1,\kappa}^{i,j})}\left[\ell\right]p_{k-1|k-1}^{i,j,\alpha}\left(y^{1:\ell};\kappa\right)dy^{1:\ell}.\label{eq:Single_branch_prediction_appendix_m_1}
\end{align}

If we evaluate (\ref{eq:Single_branch_prediction_appendix_m_1}) at
$B=\left(\left(\overline{\omega}_{k-1,\kappa}^{i,j},0\right),x^{1:\ell(\overline{\omega}_{k-1,\kappa}^{i,j})}\right)$
with $\kappa<k-1$, we obtain the first entries in (\ref{eq:prediction_density_m1_prop})
and (\ref{eq:prediction_beta_m1_prop}). If we evaluate (\ref{eq:Single_branch_prediction_appendix_m_1})
at $B=\left(\left(\overline{\omega}_{k-1,\kappa}^{i,j},0\right),x^{1:\ell(\overline{\omega}_{k-1,\kappa}^{i,j})}\right)$
with $\kappa=k-1$, we obtain the second entries in (\ref{eq:prediction_density_m1_prop})
and (\ref{eq:prediction_beta_m1_prop}). If we evaluate (\ref{eq:Single_branch_prediction_appendix_m_1})
at $B=\left(\left(\overline{\omega}_{k-1,\kappa}^{i,j},1\right),x^{1:\ell(\overline{\omega}_{k-1,\kappa}^{i,j})+1}\right)$,
we obtain the third entries in (\ref{eq:prediction_density_m1_prop})
and (\ref{eq:prediction_beta_m1_prop}). If we evaluate (\ref{eq:Single_branch_prediction_appendix_m_1})
at any other $B$, the output is zero. This finishes the proof of
(\ref{eq:prediction_density_m1_prop}) and (\ref{eq:prediction_beta_m1_prop}).

\subsection{Case $m>1$}

The probability of existence of $p_{k|k-1}^{i,(j,m),\alpha}\left(\cdot\right)$,
$m>1$, in (\ref{eq:Prediction_branch_append}) is
\begin{align}
r_{k|k-1}^{i,(j,m),\alpha} & =\int\widetilde{p}_{k|k-1}^{i,(j,m),\alpha}\left(\left\{ B\right\} \right)dB\nonumber \\
 & =\int\int g_{m}\left(\left\{ B\right\} \left|t_{k-1},\left\{ B'\right\} \right.\right)\nonumber \\
 & \,\times r_{k-1|k-1}^{i,j,\alpha}p_{k-1|k-1}^{i,j,\alpha}\left(B'\right)dB'dB.
\end{align}
As $g_{m}\left(\cdot|\cdot\right)$ requires an existing branch at
time step $k-1$ and only depends on the last state of the parent
branch, we can use (\ref{eq:single_branch_density}) to obtain
\begin{align}
r_{k|k-1}^{i,(j,m),\alpha} & =r_{k-1|k-1}^{i,j,\alpha}\beta_{k-1|k-1}^{i,j,\alpha}\left(k-1\right)\nonumber \\
 & \:\times\left\langle p_{m}^{S}\left(x^{\ell}\right),p_{k-1|k-1}^{i,j,\alpha}\left(x^{\ell};k-1\right)\right\rangle .
\end{align}

The predicted single branch density evaluated at branch $\left(\left(\overline{\omega}_{k-1,k-1}^{i,j},m\right),y\right)$
is
\begin{alignat}{1}
 & p_{k|k-1}^{i,(j,m),\alpha}\left(\left(\overline{\omega}_{k-1,k-1}^{i,j},m\right),y\right)\nonumber \\
 & =\frac{p_{k|k-1}^{i,(j,m),\alpha}\left(\left\{ \left(\overline{\omega}_{k-1,k-1}^{i,j},m\right),y\right\} \right)}{r_{k|k-1}^{i,(j,m),\alpha}}\\
 & =\frac{1}{r_{k|k-1}^{i,(j,m),\alpha}}\int g_{m}\left(\left\{ \left(\overline{\omega}_{k-1,k-1}^{i,j},m\right),y\right\} \left|t_{k-1},\left\{ B'\right\} \right.\right)\nonumber \\
 & \:\times r_{k-1|k-1}^{i,j,\alpha}p_{k-1|k-1}^{i,j,\alpha}\left(B'\right)dB'\nonumber \\
 & =\frac{1}{\left\langle p_{m}^{S}\left(x\right),p_{k-1|k-1}^{i,j,\alpha}\left(x;k-1\right)\right\rangle }\int p_{m}^{S}\left(x^{\ell}\right)\nonumber \\
 & \:\times g_{m}\left(y|x^{\ell}\right)p_{k-1|k-1}^{i,j,\alpha}\left(x^{\ell};k-1\right)dx^{\ell}\label{eq:AppendixC_equation}
\end{alignat}
where $p_{k-1|k-1}^{i,j,\alpha}\left(x^{\ell};k-1\right)$ denotes
the marginal distribution of $p_{k-1|k-1}^{i,j,\alpha}\left(\cdot;k-1\right)$
at the last time step of the branch. Eq. (\ref{eq:AppendixC_equation})
corresponds to a density of the form (\ref{eq:single_branch_density})
with one mixture component, $\beta_{k|k-1}^{i,(j,m),1}(k)=1$ and
$p_{k|k-1}^{i,(j,m),\alpha}\left(y;k\right)$ given by (\ref{eq:prediction_density_m_higher1_prop}),
completing the proof of Proposition \ref{prop:Bernoulli_tree_prediction}.

\section{\label{sec:AppendixD}}

For completeness, this appendix reviews the main aspects of the linear
programming (LP) metric for sets of trajectories. Full details are
provided in \cite{Angel20_d}. 

\subsection{Preliminary concepts}

We consider two sets of trajectories $\mathbf{T}^{x}=\left\{ T_{1}^{x},...,T_{n_{x}}^{x}\right\} $
and $\mathbf{T}^{y}=\left\{ T_{1}^{y},...,T_{n_{y}}^{y}\right\} $,
with trajectories up to a time step $k_{\mathrm{max}}$. Let $\mathbf{x}_{i}^{k}$
and $\mathbf{y}_{j}^{k}$ denote the sets of targets at time step
$k$ corresponding to trajectories $T_{i}^{x}$ and $T_{j}^{y}$.
It is met that $|\mathbf{x}_{i}^{k}|\leq1$ and $|\mathbf{y}_{j}^{k}|\leq1$,
i.e., a trajectory $T_{i}^{x}$ may not exist at time step $k$ ($\mathbf{x}_{i}^{k}=\emptyset$)
or contain a single target ($\mathbf{x}_{i}^{k}=\left\{ x\right\} $). 

Let $d_{G}\left(\cdot,\cdot\right)$ denote the generalised optimal
subpattern assignment (GOSPA) metric (with its parameter $\alpha=2$)
for sets of targets \cite{Rahmathullah17}. For sets with at most
one target, such as $\mathbf{x}_{i}^{k}$ and $\mathbf{y}_{j}^{k}$,
the GOSPA metric becomes
\begin{align}
d_{G}\left(\mathbf{x}_{i}^{k},\mathbf{y}_{j}^{k}\right) & =\begin{cases}
\min\left(c,d_{b}\left(x,y\right)\right) & \mathbf{x}_{i}^{k}=\left\{ x\right\} ,\mathbf{y}_{j}^{k}=\left\{ y\right\} \\
0 & \mathbf{x}_{i}^{k}=\mathbf{y}_{j}^{k}=\emptyset\\
\frac{c}{2^{1/p}} & \mathrm{otherwise}
\end{cases}\label{eq:GOSPA_set_1target}
\end{align}
where $d_{b}\left(\cdot,\cdot\right)$ is a base distance for single
targets, $p$ is a scalar such that \textit{$1\leq p<\infty$} and
parameter $c>0$ represents the maximum localisation error to regard
a target $x$ as being properly detected by an estimate $y$\textit{.}

We define an $(n_{x}+1)\times(n_{y}+1)$ matrix $D_{\mathbf{T}^{x},\mathbf{T}^{y}}^{k}$
whose $(i,j)$ element is
\begin{align}
D_{\mathbf{T}^{x},\mathbf{T}^{y}}^{k}(i,j) & =d_{G}\left(\mathbf{x}_{i}^{k},\mathbf{y}_{j}^{k}\right)^{p}
\end{align}
with $\mathbf{x}_{n_{x}+1}^{k}=\emptyset$ and $\mathbf{y}_{n_{y}+1}^{k}=\emptyset$.
That is, $D_{\mathbf{T}^{x},\mathbf{T}^{y}}^{k}$ contains all possible
GOSPA errors (to the $p$-th power) between all possible associations
between $\mathbf{x}_{i}^{k}$ and $\mathbf{y}_{j}^{k}$. 

In the trajectory metric, at each time step, we assign trajectories
in $\mathbf{T}^{x}$ to trajectories in $\mathbf{T}^{y}$, or leave
the trajectories unassigned. We represent these assignments using
a binary matrix $W^{k}$ that satisfies the following properties:
\begin{align}
\sum_{i=1}^{n_{x}+1}W^{k}(i,j) & =1,\ j=1,\ldots,n_{y}\label{eq:binary_constraint1}\\
\sum_{j=1}^{n_{y}+1}W^{k}(i,j) & =1,\ i=1,\ldots,n_{x}\\
W^{k}(n_{x}+1,n_{y}+1) & =0,\label{eq:binary_constraint3}\\
W^{k}(i,j) & \in\{0,1\},\forall\ i,j\label{eq:binary_constraint4}
\end{align}
where $W^{k}(i,j)$ is the element in the row $i$ and column $j$
of matrix $W^{k}$. We have $W^{k}(i,j)=1$ if $\mathbf{x}_{i}^{k}$
is associated to $\mathbf{y}_{j}^{k}$, $W^{k}(i,n_{y}+1)=1$ if $\mathbf{x}_{i}^{k}$
is unassigned, and $W^{k}(n_{x}+1,j)=1$ if $\mathbf{y}_{j}^{k}$
is unassigned. 

\subsection{LP trajectory metric}

To enable a fast computation, in the LP trajectory metric, we consider
soft assignments between the sets of targets at each time step. That
is, we consider a matrix $W^{k}\in\mathcal{\overline{W}}_{\mathbf{T}_{x},\mathbf{\mathbf{T}_{y}}}$
that meets (\ref{eq:binary_constraint1})-(\ref{eq:binary_constraint3})
and $W^{k}(i,j)\in\left[0,1\right]$. 
\begin{defn}
For $1\leq p<\infty$, cut-off $c>0$, switching penalty $\gamma>0$,
base metric $d_{b}\left(\cdot,\cdot\right)$, the LP metric for sets
of trajectories $\mathbf{T}^{x}$ and $\mathbf{T}^{y}$ is
\begin{align}
d\left(\mathbf{T}^{x},\mathbf{T}^{y}\right) & =\min_{\substack{W^{k}\in\mathcal{\overline{W}}_{\mathbf{T}^{x},\mathbf{T}^{y}}\\
k=1,\ldots,k_{\mathrm{max}}
}
}\Bigg(\sum_{k=1}^{k_{\mathrm{max}}}\mathrm{tr}\big[\big(D_{\mathbf{T}^{x},\mathbf{T}^{y}}^{k}\big)^{T}W^{k}\big]\nonumber \\
 & \quad+\frac{\gamma^{p}}{2}\sum_{k=1}^{k_{\mathrm{max}}-1}\sum_{i=1}^{n_{x}}\sum_{j=1}^{n_{y}}|W^{k}(i,j)-W^{k+1}(i,j)|\Bigg)^{\frac{1}{p}}.\label{eq:LP_metric}
\end{align}
\end{defn}
Namely, the LP trajectory metric uses the optimal soft-assignment
between trajectories in $\mathbf{T}^{x}$ and $\mathbf{T}^{y}$ at
each time step. The first term in (\ref{eq:LP_metric}) considers
the costs for localisation errors, missed and false targets. The second
term in (\ref{eq:LP_metric}) is the track switching cost. All details
on the metric decomposition into these costs are provided in \cite[Sec. IV.C]{Angel20_d}.
\end{document}